\documentclass[twocolumn]{aastex62}
\usepackage{CJK}
\usepackage{natbib}
\usepackage{graphicx}
\usepackage {amsmath,amssymb,eqnarray}
\usepackage{enumerate}
\usepackage{enumitem}
\usepackage{mathtools}

\newcommand{\hi}{{\rm H}{\textsc i}}

\begin{document}

\title{ Ram pressure stripping of HI-rich galaxies infalling into massive clusters }

\author[0000-0002-6593-8820]{Jing Wang}
\affiliation{ Kavli Institute for Astronomy and Astrophysics, Peking University, Beijing 100871, China}
\author{Weiwei Xu}
\affiliation{ Kavli Institute for Astronomy and Astrophysics, Peking University, Beijing 100871, China}
\author{Bumhyun Lee}
\affiliation{ Kavli Institute for Astronomy and Astrophysics, Peking University, Beijing 100871, China}
\author{Min Du}
\affiliation{ Kavli Institute for Astronomy and Astrophysics, Peking University, Beijing 100871, China}
\author{Roderik Overzier}
\affiliation{Observat\'{o}rio Nacional/MCTIC, Rua General Jos\'{e} Cristino, 77, S\~{a}o Crist\'{o}v\~{a}o, Rio de Janeiro, RJ 20921-400, Brazil}
\author[0000-0003-2015-777X]{Li Shao}
\affiliation{National Astronomical Observatories, Chinese Academy of Sciences, 20A Datun Road, Chaoyang District, Beijing, China}


\begin{abstract}
We estimate the strength of ram pressure stripping (RPS) for $\hi$-rich galaxies in X-ray detected clusters. We find that galaxies under stronger RPS tend to show more significantly reduced total $\hi$ mass and enhanced central SFR, compared to control galaxies in the field which have similar stellar mass, stellar surface density and integral star formation rate. Galaxies under strong or weak RPS account for $\sim$40\% of the $\hi$-rich population at $R_{200}$, and even beyond $R_{200}$ in the most massive clusters. Our results imply the important role of RPS as a channel of environmental processing far before the galaxies reach the core region of clusters.

\end{abstract}


\section{Introduction}
\label{sec:introduction}
All galaxies including the Milky Way suffer from gravitational and hydro-dynamic effects of the external environment. Strong environmental effects remove the hot and cold gases as well as stellar outer disks, leading to the cessation of star formation and aging of stellar population. With little input of the dynamically cold, young stars, disks suffering from the continuous heating from the external perturbations, undergo morphological transformation. Galaxy evolution is hence accelerated by these effects. Massive clusters provide an ideal laboratory to study a variety of environmental effects, particularly the ram pressure stripping (RPS) which is much weaker in smaller groups.  

\citet{Gunn72} provided the first analytical model of RPS, in which the interstellar gas of a galaxy gets stripped if the ram pressure is higher than the anchor force. In this model, the ram pressure depends on the density of the intra-cluster medium (ICM) and the velocity of the galaxy, and the anchor force depends the surface density of the ISM and the internal gravity on it in the direction perpendicular to the disk. Later hydrodynamic simulations found that RPS is stronger for face-on infalling galaxies than for edge-on ones \citep{Jachym09, Roediger06}, and the orbits of infall affect the cumulative effect of RPS \citep{Tonnesen19}. Pressures from the hot gas halo help constrain the interstellar gas against the ram pressure \citep{Cora18, Stevens17}, though a significant fraction of the galaxies should have lost the hot gas halo and start strangulation \citep{Larson80} during an early stage of the infall \citep{Bahe13, Bekki09}. Some of the interstellar gases stripped off the disk plane may be accreted later if they do not reach the escape velocity \citep{Vollmer01}. 
Despite these uncertainties, the model of \citet{Gunn72} is a good approximation to quantify the strength of ram pressure stripping. It is widely applied in semi-analytical models (SAM) of galaxy evolution with RPS \citep{Lotz19, Cora18, Stevens17, Luo16, Henriques15, Gonzalez-Perez14} and used to interpret observational trends. Based on the \citet{Gunn72} model, we expect RPS to be most effective in the core region of massive clusters and on the low-mass galaxies. 

Observations confirmed RPS as one effective mechanism in the evolution of both individual galaxies and the general star-forming population near the cores of clusters. Galaxies under RPS were identified in nearby clusters, with a truncated edge on the leading side and a tail on the trailing side of infalling gas disks \citep[e.g.][]{Gavazzi18, Boselli16, Yagi17, Abramson11, Chung09}. These galaxies typically show high $\hi$ deficiency, indicative of a recent gas removal and future quenching of star formation \citep{Boselli16, Chung09, Bravo-Alfaro00, Cayatte94}. They represent a subset of the galaxies under strong RPS, as the observability of tails can depend on both the observing angle and the thermal pressure of the ICM \citep{Tonnesen10}. The averaged
behavior of galaxies in clusters also supports the importance of RPS.
Both the gas mass and SFR of galaxies at a given stellar mass decrease on average toward the cluster centers \citep{Brown17, Odekon16, Hess13, Woo13}. These trends are more prominent for low-mass galaxies than for high-mass galaxies \citep{Cortese11, Zhang13}, in more massive halos than in less massive halos \citep{Brown17, Odekon16}. The interstellar medium and SFR show an averaged behavior of outside-in shrinking within galaxies near the core of clusters: when compared to the field galaxies, the most extended component $\hi$ shows the strongest deficiency, the less extended dust, molecular gas and integral SFR are also reduced but to a less extent, and the inner most central SFR is the least affected \citep{Boselli20, Mok17, Boselli14, Cortese10, Boselli06, Crowl05}. 
These trends are qualitatively consistent with the way that RPS is predicted to work, and indicate the dominating role of RPS near the core of clusters.

The relative importance of RPS among many other environmental effects is less clear in the outer region (i.e. near and beyond the virial radius) of clusters \citep{Koopmann04}. Galactic tidal stripping and mergers are expected and observed to be relatively frequent there \citep{Chung09}, as the galaxy densities are higher than in the field but the relative velocities of galaxies are not as high as those near the cluster cores \citep{Boselli06}. Strangulation due to removal of the hot gas halo is also expected to happen at much larger cluster-centric distance than RPS of the cold gas \citep{Bahe13, McCarthy08, Larson80}. 
There is still not consensus on how the star formation activity of galaxies is reduced through the clusters. A slow$+$fast declining mode is derived by several authors \citep{Wijesinghe12, Wetzel12, Wetzel13, Muzzin12}; others rather indicate a slow decline \citep{vonderLinden10, Wolf09, McGee09, Paccagnella16}, while others suggest a rapid
quenching \citep{Boselli16, Oman16}. These debates, while may be partly attributed to biases in sample selection or analysis methods, suggest a complexity in environmental effects. The galaxies with a fast declining SFR are usually associated with strong RPS, which theoretically can remove 70\% of the cold gas in a few hundreds Myrs \citep{Yun19}; while those with a slowly declining SFR may be under a mixture of effects including weak RPS. 
Some insights about the relative role of RPS could be gained from cosmological simulations, but due to the complex nature of galaxies, properly modeling the $\hi$ and SFR of central and satellite galaxies in clusters has been difficult \citep{Stevens19, Lotz19, Cora18, Stevens17, Luo16, Henriques15, Gonzalez-Perez14}. More observational inputs may help constrain the simulations.

One way of better separating the effect of RPS in observations, is to derive model motivated parameters. The ram pressure and the anchor force in the \citet{Gunn72} model can be roughly estimated from observations in X-ray, optical, and $\hi$ or ionized gas. Studies based on the IFU survey of Jellyfish galaxies, GASP \citep{Poggianti17}, found that the observed significance of RPS tails in the ionized gas are consistent with the levels of ram pressures in comparison to anchor forces \citep{Jaffe18}. Another useful tool is the projected phase-space diagram (PSD), a plot of the radial velocities as a function of the projected cluster centric distances, which effectively traces the infall stage of galaxies \citep{Mahajan11, Oman13, Oman16, Rhee17}. By requiring the ram pressure to be larger than the anchor force at all galactic radii, a ``stripping region'' of $\hi$, where galaxies are undetected in shallow $\hi$ surveys, was successfully predicted on the projected PSD for several massive clusters \citep{Jaffe15, Jaffe16, Yoon17}. Studies based on
high-resolution $\hi$ images show $\hi$ richness consistent with infall stages indicated by the projected PSD positions, but the galaxies displaying RPS tails are typically found beyond the stripping regions \citep{Yoon17}, which is reasonable as the RPS timescale in the stripping region is short (a few tens of Myr, \citealt{Abadi99}).

$\hi$ is an excellent tracer of the relatively early stage of environmental processing on star-forming galaxies. It is the reservoir for forming stars, and sensitive to perturbations when it is more extended than the stellar disks. Thus, its richness is associated with both the strength of environmental effects and the progress of star formation cessation \citep{Boselli06, Boselli14b}. High-resolution interferometric images for selected galaxies in nearby clusters, particularly in Virgo, have revealed the morphological and kinematic features of $\hi$ in galaxies in response to RPS and other environmental processes \citep{Chung09, Yoon17}. They provide valuable constraints to zoom-in simulations modeling physical details of the RPS process (e.g. \citealt{Tonnesen09}). On the hand, low-resolution (usually single-dish) but blind and contiguous $\hi$ surveys provide opportunities to completely map a cluster out to several times the virial radius, and cover statistically significant number of clusters and galaxies \citep{Haynes18, Jaffe16, Jaffe15}. Studies based on this type of data characterize the statistical behavior of galaxies when they are potentially affected by RPS \citep{Odekon16, Yoon15, Hess13}. These statistical observational results help constrain simulations under a cosmological context, focusing on  the role that RPS plays among many other processes in the general evolution of galaxies (e.g. \citealt{Stevens19, Yun19}). 

A major limitation of the  blind $\hi$ surveys is the low resolution, while RPS is predicted to be dependent on the $\hi$ surface density at a given radius in the galaxies \citep{Gunn72}. Hence, predicting the $\hi$ radial distribution based on rules extracted from the interferometry data of nearby galaxies will provide some insights when one attempts to link the observed total $\hi$ mass to the RPS process. Such an analysis has been often used in statistical studies with low-resolution $\hi$ data on the topic of RPS \citep{Boselli18, Jaffe16, Jaffe15, Boselli14a} \footnote{The gas is assumed to be distributed exponentially in the multi-zone chemo spectrophotometric model of  \citet{Boselli18,Boselli14a}. }. Exploring methods to enhance the science value of low-resolution $\hi$ data is also in line with a preparation for SKA pathfinder $\hi$ surveys, for the number of new $\hi$ detections will explode but the majority of them will be unresolved in these surveys \citep{StaveleySmith15}. 

Recent advances in observations enable us to predict $\hi$ radial profiles with higher accuracy than before. 
Based on $\hi$ images for over 500 nearby galaxies, \citet{Wang16} found that all these galaxies lie tightly on a relation between the $\hi$ mass and a characteristic radius of the $\hi$ disks ($R_{\rm HI}$); the relation is partly because the outer region of $\hi$ disks have similar radial profiles when the radius is normalized by $R_{\rm HI}$ (also see \citealt{Wang14}).
 The similarities seem to be a result from the sophisticated balance between different physical processes, including the accretion, radial flow, and depletion of the $\hi$ \citep{Bahe16, Wang14}. Using the median $\hi$ radial profile (normalized by $R_{\rm HI}$) from \citet{Wang16}, \citet{Wang20} showed that the $\hi$ mass beyond and within the optical radius of galaxies can be predicted to a high accuracy. Because the majority of the galaxies from the Virgo cluster lie on the same $\hi$ size-mass relation, and exhibit a similar median $\hi$ radial profile as the field galaxies \citep{Wang16}, we may  also use these characteristics to predict the $\hi$ radial profile of galaxies in clusters for statistical studies of RPS. 

In this paper, we combine the \citet{Gunn72} model with the projected PSD, and conduct a statistical analysis of RPS effects on $\hi$-detected galaxies beyond the stripping regions of X-ray detected, massive clusters. We modify the classical way of estimating the anchor forces, by better predicting the $\hi$ distribution in individual galaxies. Compared to many earlier studies which characterized the relatively strong type of RPS with galaxies showing significant $\hi$ deficiencies, we more focus on a relatively early stage of RPS which is weak and has not strongly depleted the $\hi$ or suppressed the SFR of galaxies yet. We attempt to evaluate the statistical significance of weak RPS among the HI-rich population falling into clusters, particularly in the cluster outer regions where many environmental effects co-exist.
We ask out to what cluster centric radius does RPS occur in the selected clusters? What is the fraction of $\hi$-rich galaxies affected by RPS at each cluster centric distance? How different are the galaxies under relatively weak RPS from the field galaxies with similar integral SFR?
We present the sample selection in Sec.~\ref{sec:data}, and the estimate of RPS strengths in Sec~\ref{sec:analysis}. We present the results in Sec.~\ref{sec:results}, discuss in Sec.~\ref{sec:discussion} , and conclude in Sec.~\ref{sec:conclusion}. Throughout this paper, we assume a Chabrier initial mass function \citep{Chabrier03}, and a $\Lambda$CDM cosmology with $\Omega_{m}=0.3$, $\Omega_{\lambda}=0.7$ and $h=0.7$. We do not account for the contribution of helium when discussing $\hi$ properties, unless specified.

\section{Data}
\subsection{X-ray detected clusters}
\label{sec:data}
\subsubsection{The HIFLUGCS and RXGC samples}
\label{sec:data_cluster}
We use two X-ray samples of nearby clusters, the HIghest X-ray FLUx Galaxy Cluster Sample (HIFLUGCS, \citealt{Reiprich02}) and RASS-based extended X-ray Galaxy Cluster Catalog (RXGCC, Xu et al. in prep). 

HIFLUGCS selected from the ROSAT All-Sky Survey (RASS) the 63 brightest clusters, with Galactic latitude $|b_{II}|>20^{\circ}$, and outside the LMC, SMC and Virgo regions.  
\citet{Reiprich02} fit beta models to the X-ray surface brightness profiles of the high-mass clusters
\begin{equation}
S(r)=S(0)(1+r^2/r_c^2)^{-3\beta+1/2},
\end{equation}
where $S(r)$ is the surface brightness at radius $r$. Assuming the ICM to be in hydrostatic equilibrium and isothermal, they
derived $R_{200}$ based on the best-fit X-ray surface brightness models, where $R_{200}$ is the radius within which the averaged mass density is 200 times the critical cosmic matter density at the redshift. Their assumption of ICM status ignored the local dynamics, but provides reasonable description for large-scale properties like $R_{200}$.
 The HIFLUGCS clusters have been extensively studied in the literature, including their detailed ICM distributions \citep{Eckert11}.

RXGCC used a state-of-art algorithm, which includes the wavelet filtering, source extraction and maximum likelihood fitting,
 to extract from X-ray images the low surface density groups or clusters, and built a sample of 764 clusters and groups from RASS (\citealt{Xu18}, Xu et al. in prep). In Xu et al. 2018, the R$_{500}$ is obtained from the appearance of galaxy clusters and the significance radius, which is derived from the growth-curve analysis. They further derived $R_{200}\approx1.538R_{500}$ assuming NFW profiles with a concentration index of 4. 
 The RXGCC optimized the measurements for the faint clusters and groups, and is complementary to the HIFLUGCS sample. 
 
We select clusters with redshift $z<0.05$ to match the ALFALFA redshift range, and are left with 45 and 207 clusters in HIFLUGCS and RXGCC.
The $M_{200}$ distribution of HIFLUGCS after the redshift selection has 20, 50 and 80 percentiles of 1.63, 4.52 and 6.73$\times10^{14}~M_{\odot}$, and for RXGCC the percentiles are 0.42, 0.89 and 1.84 $\times10^{14}~M_{\odot}$. We select clusters with $M_{200}>2\times 10^{14} ~ M_{\odot}$ and $<2\times 10^{14} ~ M_{\odot}$ from HIFLUGCS and RXGCC respectively, and are left with 37 and 170 clusters in each sample. We call the subsamples out of HIFLUCGS and RXGCC the high-mass and low-mass cluster samples respectively in the following analysis. Despite the relative difference in mass, both types of clusters are massive clusters compared to optically selected groups, as suggested by the detection in the X-ray.

\subsubsection{ Parameters that define the cluster region }
In the next section, we select member galaxies of the clusters based on the distribution of galaxies on the projected PSD of radial velocity difference as a function of  projected cluster centric distance.

We firstly derive the relation of the escape velocity ($v_{esc,0}$) as a function cluster centric distance ($d$), assuming an Navarro-Frenk \& White (NFW, \citealt{Navarro97}) profile for the dark matter distribution. We assume for the NFW profiles a concentration index of 4. Because of projection effects, we replace $d$ with the averaged, projected cluster centric distance $d_{proj}\sim \pi/4 d$, and because only radial velocities are observable, we replace $v_{esc}$ with the averaged, radial escape velocity $v_{esc} \sim v_{esc,0}/\sqrt{3}$. The $v_{esc}$-$d_{proj}$ relation has been proven to be effective in identifying infalling and settled galaxies of clusters \citep{Oman13}.

In this paper, a galaxy is identified as a (settled or infalling) member of a cluster if $d_{proj}<2R_{200}$, and the radial velocity difference $|\Delta v_{rad}|$ is smaller than the value of $v_{esc}$ expected at $d_{proj}$.


\subsection{HI detected galaxies}
\label{sec:data_galaxy}
\subsubsection{ Selection from ALFALFA, MPA$/$JHU catalog, and GSWLC-2 }
ALFALFA \citep{Haynes18} mapped 7000 deg$^2$ in the southern sky with a typical rms of 1.6 mJy at a channel width of 18 km$/$s (after smoothing). The angular resolution (beam size) is 3.5 arcmin in full-width half-maximum. ALFALFA provided each detected galaxy the integral $\hi$ mass, $M_{\rm HI}$.
We estimate the characteristic radius $R_{\rm HI}$, based on the $\hi$ size-mass relation \citep{Wang16}, where $R_{\rm HI}$ is the semi-major axis of the 1$M_{\odot}~pc^{-2}$ isophote of $\hi$ disks.  The ALFALFA catalog has assigned each $\hi$ detection an optical counterpart from SDSS or DSS, based on the projected distance, redshift, color, morphology, and additional scientific judgement from its authors \citep{Haynes11}. We will use the optical coordinates of the ALFALFA detected galaxies in cross-matching with other galaxy catalogs. 

The MPA$/$JHU catalog \citep{Kauffmann03} provides spectroscopic measurements of galaxies from Data Release 7 of SDSS \citep{Abazajian09}. We will use the spectral indices D$_{4000}$, and equivalent width of the H$\alpha$ emission ($EW(H\alpha)$, positive values for emissions here) to indicate the star forming status in the galactic center. D$_{4000}$ is produced mainly because spectrum to the blue side of 4000 $\AA$ is strongly absorbed by metals in the atmosphere of old stars. So low values of D$_{4000}$ suggest existence of a significant amount of young stellar population. $H\alpha$ emission is produced by ionizing radiation from O stars with a typical life time of 10 Myr. 
We also use the photometric measurements of $g$-band $R_{25}$ (semi-major axis of the 25 mag arcsec$^{-2}$ isophote), $i$-band $R_{50}$ (half-light radius), and $r$-band $R_{90}$ ($90\%$-light radius), $R_{50}$ and $R_d$ (the scale-length). We calculate $\mu_*$, the averaged stellar surface density within the $i$-band $R_{50}$, and the concentration index $R_{90}/R_{50}$ in the $r$-band.

The GALEX-SDSS-WISE Legacy Catalog 2 (GSWLC-2, \citealt{Salim16}) selected galaxies overlapping in GALEX \citep{Martin05}, SDSS and WISE \citep{Wright10}. It estimated the integral star formation rate (SFR) and stellar mass ($M_*$) of galaxies by fitting stellar synthesis models to the broad-band spectral energy distribution ranging from the mid-infrared to the far-ultraviolet. The SFRs are mainly indicated by the attenuation corrected ultraviolet light, with sensitivity on a timescale of $\sim$100 Myr. 

We firstly search for member galaxies of each cluster from the ALFALFA catalog, using the criterial based on $R_{200}$ and $v_{esc}$. We match the optical coordinates of the AFALFA detected galaxies to the MPA$/$JHU catalog, and then to GSWLC-2, by requiring the projected distance to be less than 3 arcsec, and the radial velocity difference less than 200 km/s. These cross-matching result in 158 and 144 galaxies in high- and low-mass clusters. We note that the sample size is much smaller a few previous studies on ALFALFA and SDSS detected galaxies in X-ray clusters \citep[e.g.][]{Odekon16}, mostly because of our selection by $v_{esc}$ instead of using group catalogs built with friend-of-friend member finders \citep[e.g.][]{Yang07}. 

Motivated by the previous results that environmental effects on gas content and SFRs are most significant in low-mass galaxies \citep{Woo17, Boselli14a, Boselli14b, Catinella13, Wetzel13, Gavazzi10}, we select the galaxies with $\log M_*/M_{\odot}<11$. In order to focus on the $\hi$-rich galaxies under active environmental processing instead of already being fully processed, we further limit the sample to galaxies with $R_{\rm HI}>R_{25}$. By this selection, we focus on RPS of the $\hi$ gas and may miss galaxies which have little $\hi$ but the ionized gas still being stripped by the ram pressure (e.g. NGC 4569 in the Virgo cluster, \citealt{Chung09, Boselli18}). 

 These selection criteria reduce the sample to 142 and 128 galaxies in the high and low-mass clusters, which we refer to as the cluster sample.

\subsubsection{The control sample and final main sample}
\label{sec:data_control}
\begin{figure*} 
\includegraphics[width=18cm]{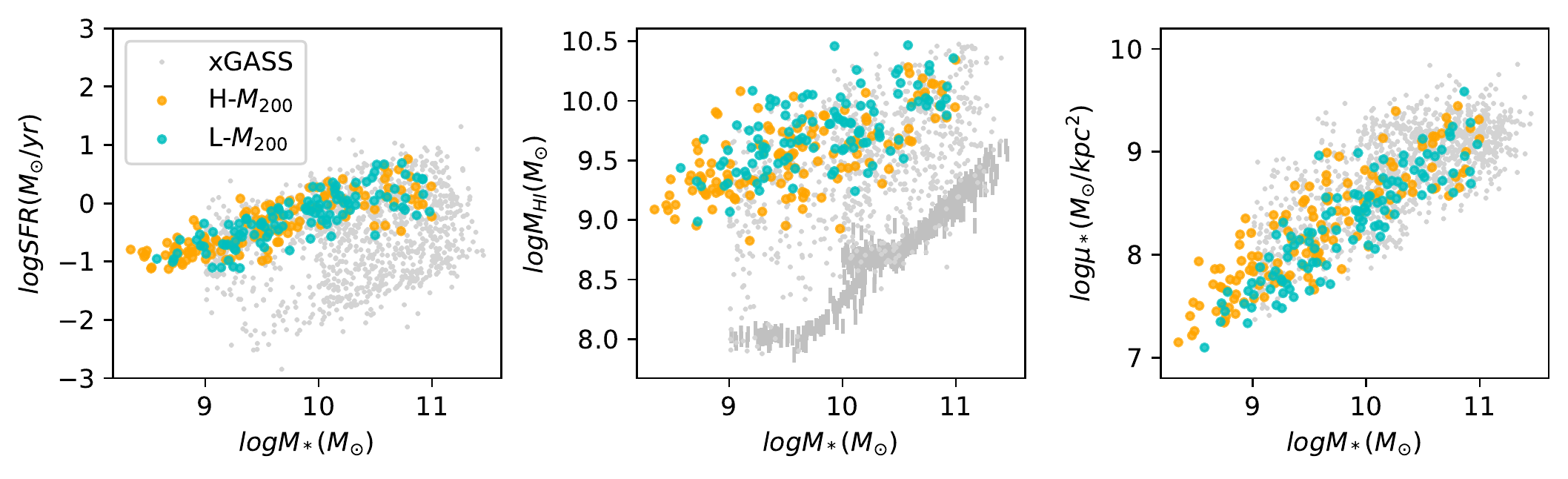}
\caption{ Scaling relations of selected cluster galaxies. From left to right, we plot the relations of SFR, $M_{\rm HI}$ and $\mu_*$ as a function of $M_*$ for galaxies in low-mass (cyan) and high-mass (orange) clusters. The xGASS \citep{Catinella18} sample of $M_*$ and $z$ selected galaxies are plotted in grey as a reference. The upper limits of $M_{\rm HI}$ in the xGASS sample are plotted as vertical bars. }
\label{fig:galprop}
\end{figure*}

We first built a pool of all galaxies with redshift below 0.05 and detected simultaneously in the ALFALFA, MPA/JHU and GWSLC-2 catalogs. It includes 13273 galaxies. For each galaxy in the cluster sample, we randomly select 8 control galaxies with replacement from the galaxy pool, with $M_*$, $SFR$, $\mu_*$ and $z$  differing by less than 0.15 dex, 0.2 dex, 0.25 dex and 0.002 respectively. We require each galaxy from the cluster sample to have at least 5 unique control galaxies, which reduces the sample to 134 and 109 galaxies for the high- and low-mass clusters. {\it This is our final, main sample of galaxies.} We show their distributions in the diagrams of SFR, $M_{\rm HI}$, and $\mu_*$ versus $M_*$ (Fig.~\ref{fig:galprop}). They are by selection strongly biased toward the $\hi$-rich, star-forming, low surface density population, and do not represent the general galaxy population in clusters which are on average older and gas poorer than the field galaxies. These galaxies are useful when searching for signatures of the relatively early and weak environmental processing prevalent in the relatively outer region of clusters, which may just start to deviate galaxies from the parameter space of unperturbed galaxies. They are on the other hand, highly incomplete when studying the relatively strong environment effects (including strong RPS) which can quick move galaxies below the detection limit of ALFALFA.

These galaxies are from 7 high-mass clusters and 19 low-mass clusters. We list properties of these clusters, including the number of selected galaxies from each cluster, in Table~\ref{tab:cluster}. 
Among the galaxies from high-mass clusters, 43\% of them come from the Coma cluster, another 49\% come from A1367, A2147 and MKW8, and the rest from the remaining 3 clusters. For low-mass clusters, the galaxies are more evenly contributed by each cluster, with a median number of 5 per cluster. 

\begin{table*}
\centering
{
\begin{tabular}{cccccccccccccc }
Name  & $N_{ga}$ &  RA &  DEC  & Redshift   &  $M_{200}$ &  $R_{200} $  & $R_{500}$  & $\sigma_C$ & $\beta$ & $N_{H}(0)$ & $N_{H,2}(0)$  & $r_c$  & $r_{c,2}$  \\

& & (deg) & (deg) & &  ($10^{14} M_{\odot}$) & (Mpc) & (Mpc) & ($km/s$)  & & ($cm^{-3}$) & ($cm^{-3}$) & (kpc) & (kpc) \\
(1) & (2) & (3) & (4) & (5) & (6) & (7) & (8) & (9) & (10) & (11) & (12) & (13) \\

\hline
\multicolumn{13}{c}{High-mass clusters} \\
A1367 &26 &  176.1903 &   19.7030 & 0.022 & 3.95 & 1.49 & 0.95 &633.9 &0.62   &  0.0011 &     - &  290 &  -\\
A2052 &6 &  229.1846 &    7.0211 & 0.035 & 2.15 & 1.22 & 0.87 &516.8 &0.75   &  0.0016 &  0.0250 &  159 &   32\\
A2063 &5 &  230.7734 &    8.6112 & 0.035 & 3.15 & 1.38 & 0.97 &587.5 &0.73   &  0.0018 &  0.0064 &  194 &   54\\
A2147 &24 &  240.5628 &   15.9586 & 0.035 & 3.36 & 1.41 & 1.00 &600.3 &0.37   &  0.0022 &     - &   61 &  -\\
COMA &58 &  194.9468 &   27.9388 & 0.023 &13.46 & 2.24 & 1.51,&956.9 &0.65   &  0.0034 &     - &  249 &  -\\
MKW3 &1 &  230.4643 &    7.7059 & 0.045 & 3.36 & 1.41 & 0.98 &600.3 &0.63   &  0.0048 &  0.0172 &   86 &   27\\
MKW8 &15 &  220.1596 &    3.4717 & 0.027 & 2.31 & 1.24 & 0.82 &529.4 &0.50   &  0.0026 &     - &   94 &  -\\
\hline
\multicolumn{13}{c}{Low-mass clusters} \\
RXG306 &6 &  129.4660 &   25.1040 & 0.028 & 1.06 & 1.02 & 0.67 &451.8 &0.66   &  0.0031 &     - &   96 &  -\\
RXG325 &1 &  138.9910 &   17.5620 & 0.029 & 0.80 & 0.93 & 0.61 &410.6 &0.66   &  0.0029 &     - &   87 &  -\\
RXG327 &4 &  139.9350 &   33.7620 & 0.023 & 0.43 & 0.75 & 0.49 &333.9 &0.66   &  0.0027 &     - &   70 &  -\\
RXG367 &1 &  155.4160 &   23.8950 & 0.040 & 1.22 & 1.09 & 0.72 &474.1 &0.66   &  0.0030 &     - &  102 &  -\\
RXG389 &2 &  162.5900 &    0.2790 & 0.041 & 0.95 & 1.01 & 0.66 &434.9 &0.66   &  0.0028 &     - &   94 &  -\\
RXG395 &6 &  164.6220 &    1.6030 & 0.039 & 0.91 & 0.99 & 0.65 &429.8 &0.66   &  0.0028 &     - &   93 &  -\\
RXG401 &11 &  167.6670 &   28.7090 & 0.032 & 0.99 & 1.01 & 0.66 &441.7 &0.66   &  0.0030 &     - &   94 &  -\\
RXG411 &1 &  169.0880 &   29.3040 & 0.047 & 1.43 & 1.16 & 0.76 &499.6 &0.66   &  0.0029 &     - &  109 &  -\\
RXG457 &7 &  181.0140 &   20.2900 & 0.023 & 0.38 & 0.72 & 0.47 &320.9 &0.66   &  0.0026 &     - &   68 &  -\\
RXG515 &7 &  200.0590 &   33.1510 & 0.037 & 0.89 & 0.98 & 0.64 &426.6 &0.66   &  0.0029 &     - &   92 &  -\\
RXG527 &7 &  202.4190 &   11.7790 & 0.023 & 0.49 & 0.78 & 0.51 &348.4 &0.66   &  0.0028 &     - &   73 &  -\\
RXG538 &9 &  203.6270 &   34.7030 & 0.025 & 0.39 & 0.73 & 0.48 &323.0 &0.66   &  0.0026 &     - &   68 &  -\\
RXG597 &6 &  223.2440 &   16.7010 & 0.044 & 1.46 & 1.17 & 0.77 &502.9 &0.66   &  0.0030 &     - &  109 &  -\\
RXG615 &3 &  228.1770 &    7.4190 & 0.046 & 1.25 & 1.11 & 0.73 &477.7 &0.66   &  0.0029 &     - &  104 &  -\\
RXG62 &3 &   18.2650 &   15.4990 & 0.043 & 1.26 & 1.11 & 0.73 &478.7 &0.66   &  0.0029 &     - &  104 &  -\\
RXG632 &4 &  233.1070 &    4.7720 & 0.039 & 1.62 & 1.20 & 0.79 &520.6 &0.66   &  0.0031 &     - &  112 &  -\\
RXG651 &6 &  241.2110 &   17.7320 & 0.035 & 1.93 & 1.26 & 0.83 &553.0 &0.66   &  0.0033 &     - &  119 &  -\\
RXG653 &3 &  241.3810 &   16.4410 & 0.042 & 0.76 & 0.94 & 0.61 &404.4 &0.66   &  0.0027 &     - &   88 &  -\\
RXG931 &7 &  333.7340 &   13.8340 & 0.026 & 0.46 & 0.77 & 0.51 &341.8 &0.66   &  0.0027 &     - &   72 &  -\\
\end{tabular}
}
\caption{ {\bf Cluster properties.} Column (1): name. Column (2): number of galaxies included in our main sample.Column (3)-(6): RA, DEC, $z$, $M_{200}$ and $R_{200}$ of the cluster centers; high-mass cluster values are taken from \citet{Reiprich02}, and low-mass cluster values from \citet{Xu18}. Column (7), (9) (12) and (13): $R_{500}$ and (double-)beta model parameters $\beta$, and $r_c$ ($r_{c,2}$); high-mass cluster values are taken from \citet{Eckert11}, and low-mass values from \citet{Xu18}. When only a single-beta model is available, the parameters for the second beta component are written as -. Column (7): velocity dispersion of the clusters, estimated from $M_{200}$ (see Sec.~\ref{sec:data_cluster} ). Column (10) and (11): central density of ICM for (double-)beta models, estimated from $M_{gas,500}$ and the (double-)beta models (see. Sec.~\ref{sec:ana_RPS}); $\rho_H=1.4N_H m_p$, where $m_p$ is the mass of proton. }
\label{tab:cluster}
\end{table*}

\section{Analysis}
\label{sec:analysis}
\subsection{ICM density and ram pressure}
\label{sec:ana_RPS}
The ram pressure stripping (RPS) strength of the ICM is calculated as $\rho~(\Delta  v)^2$, where $\rho$ is the mass density of the ICM and $\Delta v$ is the relative velocity between the ICM and the galaxy.

The density distribution of an isothermal ICM is related to the beta model of the X-ray surface brightness according to
\begin{equation}
\rho(r)=\rho(0)(1+r^2/r_c^2)^{-3\beta/2}.
\end{equation}
$r_c$ and $\beta$ are the same as the model of the X-ray surface brightness, and $\rho(0)$ can be derived by integrating the profile out to $R_{500}$, and comparing the result with the gas mass ($M_{gas,500}$) expected from scaling relations. We can use the scaling relation from \citet{Ettori15} to estimate $M_{gas,500}$ from $M_{500}$, the mass within $R_{500}$. So the key parameters needed are $R_{500}$ and the parameters of the beta model ($\beta$ and $r_c$), which are derived in different ways for the high-mass and low-mass clusters. 

Because nearly one third of the X-ray luminous, high-mass clusters have cool cores, deviating from the hydrostatic equilibrium and isothermal state, \citet{Eckert11} use a scaling relation to estimate $R_{500}$ from the virial temperature of the clusters \citep{Hudson10}. The median ratio of $R_{200}$ from  \citet{Reiprich02} over $R_{500}$ from  \citet{Eckert11} is 1.44$\pm$0.08, comparable to $R_{500}/R_{200}=1.50$ expected from an NFW profile, assuming a halo mass of 3$\times10^{14} M_{\odot}$ and a concentration of 4. 
\citet{Eckert11} combined XMM-Newton and Chandra data to derive the X-ray surface brightness profiles for the high-mass clusters. Because a single-beta model does not describe the shape of the radial profile in cool core clusters well, \citet{Eckert11} fit a double-beta model for cool core clusters, and a single beta model for No-cool core clusters. We use the (double-) beta models from \citet{Eckert11} to derive ICM densities. 

Following \citet{Xu18}, we assume a single beta model for the low-mass clusters, fixing $\beta=2/3$ and $r_c=R_{500}/7$, where $R_{500}$ has been obtained from the curve-of-growth fitting. Then we estimate $M_{gas}$ and $\rho(0)$ in a similar way as for high-mass clusters. 

We approximate the ram pressure as $P= \rho(d_{proj})~\Delta v_{rad}^2$. Such approximation has a few uncertainties, including
\begin{itemize}
\item Projection effects. $\rho(d_{proj})$ can only be viewed as an upper limit of $\rho(d)$. Similarly, $\Delta v_{rad}$ can only be viewed as a lower limit of $\Delta v$. Despite these obvious offsets, we find that ram pressure estimated in this way still leads to useful analysis. We will further discuss the influence of the projection effects on our main results in Sec.~\ref{sec:discussion}.

\item Extrapolation effects. The ROSAT data typically does not detect ICM out to $2 R_{200}$. For high-mass clusters, the typical maximum radius to detect X-ray flux in a cluster is $\sim 0.92 R_{200}$ \citep{Reiprich02}, and for low-mass clusters, it is $\sim 0.63 R_{200}$ (Xu in prep.). So when $d_{proj}$ is larger than the maximum detectable radius, $\rho(d_{proj})$ has uncertainties due to extrapolation.

\item Sub-structures in the $\rho$ distribution. These structures are typically associated with infalling groups or galactic mergers which induce shocks in the ICM \citep{Ruggiero19, Roediger14, Tonnesen08, Markevitch07}, and sometimes significantly raise the local level of ram pressure \citep{Kenney04}. This type of shocks were found in some of our selected clusters (e.g. A 1367, \citealt{Ge19}). As most of the cluster merger associated shocks found so far are distant ($>1$ Mpc) from the cluster center and weak (with Mach number $<3$, \citealp{Markevitch05, Markevitch07, Ogrean14, Itahana15, Dasadia16}), we assume that the filling factor of strong shock fronts to be small in a typical cluster at low redshift, and do not significantly affect statistical analysis. 

\item Isothermal assumption. Because the temperature drops in the core region of high-mass, cool-core clusters, $\rho$ in the same region are likely under-estimated. But because the temperature variations are typically less than twice in the core region of clusters in the high-mass cluster sample \citep{Hudson10}, and the X-ray power emissivity scales with $\rho^2T^{0.5}$ (so for the same X-ray surface density, $\rho \sim T^{-0.25}$),  the under-estimation of $\rho$ should be small ($<0.1$ dex). 

\end{itemize}

\subsection{HI density profile and anchor force}
\label{sec:ana_anchor_force}
The anchor force to hold the interstellar medium gas at a radius $r$ in the galactic disc plane can be calculated as 
\begin{equation}
F_r=2 \pi G (\Sigma_{*,r}+\Sigma_{{\rm HI},r} )\Sigma_{{\rm HI},r}  , 
\label{eq:anchor_force1}
\end{equation}
where $\Sigma_{*,r}$ and $\Sigma_{{\rm HI},r}$  are the stellar and $\hi$ surface density at $r$. This is a modified form of the \citet{Gunn72} formalism, to take into account the self-gravity of the $\hi$ gas, which cannot be ignored in the outer disks. Similar modifications can be found in \citet{Stevens17, Fujita04, Abadi99}.

We estimate the anchor forces $F_{R25}$ and $F_{RHI}$ at two characteristic radii, $R_{25}$ and $R_{\rm HI}$. 
 We use the method outlined in \citet{Wang20} to estimate $\Sigma_{{\rm HI},R25}$ for each galaxy. The method makes use of the $\hi$ size-mass relation, and the homogeneous shape of $\hi$ radial profiles in the galactic outer region. The method works best when the given radius is within the exponential dropping part of the $\hi$ surface density profile, and $R_{25}$ is a good option of such radius. Following the previous work of \citet{Jaffe18} and others, we assume an exponential disk for the stars, so that the central stellar surface density $\Sigma_{*,0}=M_*/2\pi R_d^2$, and $\Sigma_{*,r}=\Sigma_{*,0} e^{-r/R_d}$, where $R_d$ is the $r$-band scale-length. 
 
There are also a few uncertainties related to the above estimates. 
\begin{itemize}
 \item Assumption of universal $\hi$ radial profiles in the outer disks. Galaxies in clusters may have perturbed $\hi$ profiles. Luckily, \citet{Wang16} found that galaxies from the VIVA (VLA Imaging of Virgo Spirals in Atomic Gas, \citealt{Chung09}) survey lie on the same $\hi$ size-mass relation as other galaxies. We further test our method of estimating $\Sigma_{{\rm HI},R25}$ with the VIVA data. When galaxies are selected in the same $M_*$ and $R_{\rm HI}/R_{25}$  range as our main sample (in total 16 galaxies), the median offset between the predicted and observed values of $\Sigma_{{\rm HI},R25}$ is 0.08$\pm$0.16 dex (see Appendix B). In comparison, the distribution of $\Sigma_{{\rm HI},R25}$ has a scatter of 0.31 dex, so the prediction indeed helps constrain the value. 
 
 \item Only disc stars and $\hi$ gas are considered in the gravitational potential. We ignored gravity from the bulge stars, molecular gas and dark matter.  Because we are considering the anchor forces in a relatively distant outer region, the gravity from a central spheroidal bulge is usually small \citep{Abadi99}. The molecular disks usually do not extend beyond $R_{25}$ and should contribute little to the disk gravity. The contribute of the dark matter to the gravity that holds gas in the disc mid-plane should be negligible due to the low volume density, which was confirmed in \citet{Jaffe18}. 
 
 \item Over-estimates of the disk masses. By ignoring the bulge, we may over-estimate $\Sigma_{*,0}$, and therefore over-estimate the anchor force. We use the linear equation of \citet{Catinella13} which was based on the galactic decomposition catalog of \citet{Gadotti09} to roughly convert the concentration index $R_{90}/R_{50}$ to the bulge-to-total mass ratio $B/T$. We test by using the derived $1-B/T$ to scale down $\Sigma_{*,r}$ and recalculating the anchor forces. We do not find the results presented later in Sec.~\ref{sec:results} to significantly change after this treatment. This is because by selection our galaxies are disk-dominated, with $R_{90}/R_{50}$ having 10, 50 and 90 percentiles of 1.97, 2.31 and 2.73, corresponding to estimated $B/T$ of 2\%, 17\% and 36\%. However, we note that estimating $B/T$ in this way is crude, particularly because our galaxies extend to much lower $M_*$ than the limit ($>10^{10}~M_{\odot}$) of \citet{Gadotti09}. 

 \item We ignored the protection$\slash$pressure from the circum-galactic hot gas halo, which needs to be stripped before the $\hi$ gas directly feels the ram pressure. As previous studies suggested that strangulation of galaxies due to halo gas removal starts at 5$R_{200}$ from the cluster center \citep{Bahe13}, the problem of ignoring pressure from the halo gas is mitigated. 
\end{itemize}

\subsection{The strong, weak, and no-RPS galaxies}
We can view the ratio between the ram pressure and the anchor force as a measure of RPS strength. We divided our sample of galaxies into 3 types with the following criteria:
\begin{itemize}
\item {\bf Strong-RPS:} $P> F_{R25}$, including 17 (13\%) and 3 (3\%) galaxies in high-mass and low-mass clusters, respectively. 
\item {\bf Weak-RPS: }  $F_{RHI}<P< F_{R25}$, including 70 (52\%) and 33 (30\%) galaxies in high-mass and low-mass clusters, respectively. 
\item {\bf No-RPS:}  $P<F_{RHI}$, including 48 (35\%) and 73 (67\%) galaxies in high-mass and low-mass clusters, respectively. 
\end{itemize}

Atlases of SDSS images and ALFALFA $\hi$ spectrum of the 20 strong-RPS galaxies can be found in Sec.~\ref{sec:appendix_atlas} and \ref{sec:appendix_spec} in the Appendix.

By definition we expect that on average the strong-RPS galaxies are strongly stripped by ram pressure near or on the stellar disks, the weak-RPS galaxies just start to feel the ram pressure in the very outer region of their $\hi$ disks, and no-RPS galaxies are not strongly affected by ram pressure in either the $\hi$ or stellar disks. 

Because both ram pressure and anchor force have uncertainties in the estimates, the division into these three groups should be viewed as being statistically representative instead of being absolutely accurate. We also remind that all the galaxies in the main sample are ALFALFA detected, $\hi$-rich galaxies, and do not represent the general galaxy population. The sample should have also significantly missed the galaxies which suffer from strong RPS and become highly deficient in $\hi$. The galaxies in this sample are thus selected to search for candidates under relatively weak RPS on the $\hi$ and at a relatively early stage of being processed by the environment, when they are moving through the clusters.  
 
\section{Results}
\label{sec:results}
 \subsection{Projected PSD positions versus internal surface densities and the determination of RPS strength}
 Previous studies have demonstrated that the projected PSD is not only reliable for identifying cluster members, but also useful to statistically assign cluster members to different infall stages \citep{Oman13}. Virialized galaxies tend to lie in a triangular region around the cluster center in the projected PSD, while galaxies that are infalling for the first time tend to have higher $\Delta v$ at a given $d_{proj}$. For convenience, we call them virialized and infalling galaxies respectively. We keep in mind that galaxies may transport between the pericentre and apocentre (the ``back-splash''  galaxies) several times before finally virialized. According to the literature, nearly half of the galaxies at small $|\Delta v_{rad}|$ (e.g. $<\sigma_C$) and slightly beyond the virialized region (1-$2R_{200}$) can be back-splash objects \citep{Mahajan11}. 

\begin{figure*} 
\includegraphics[width=8.5cm]{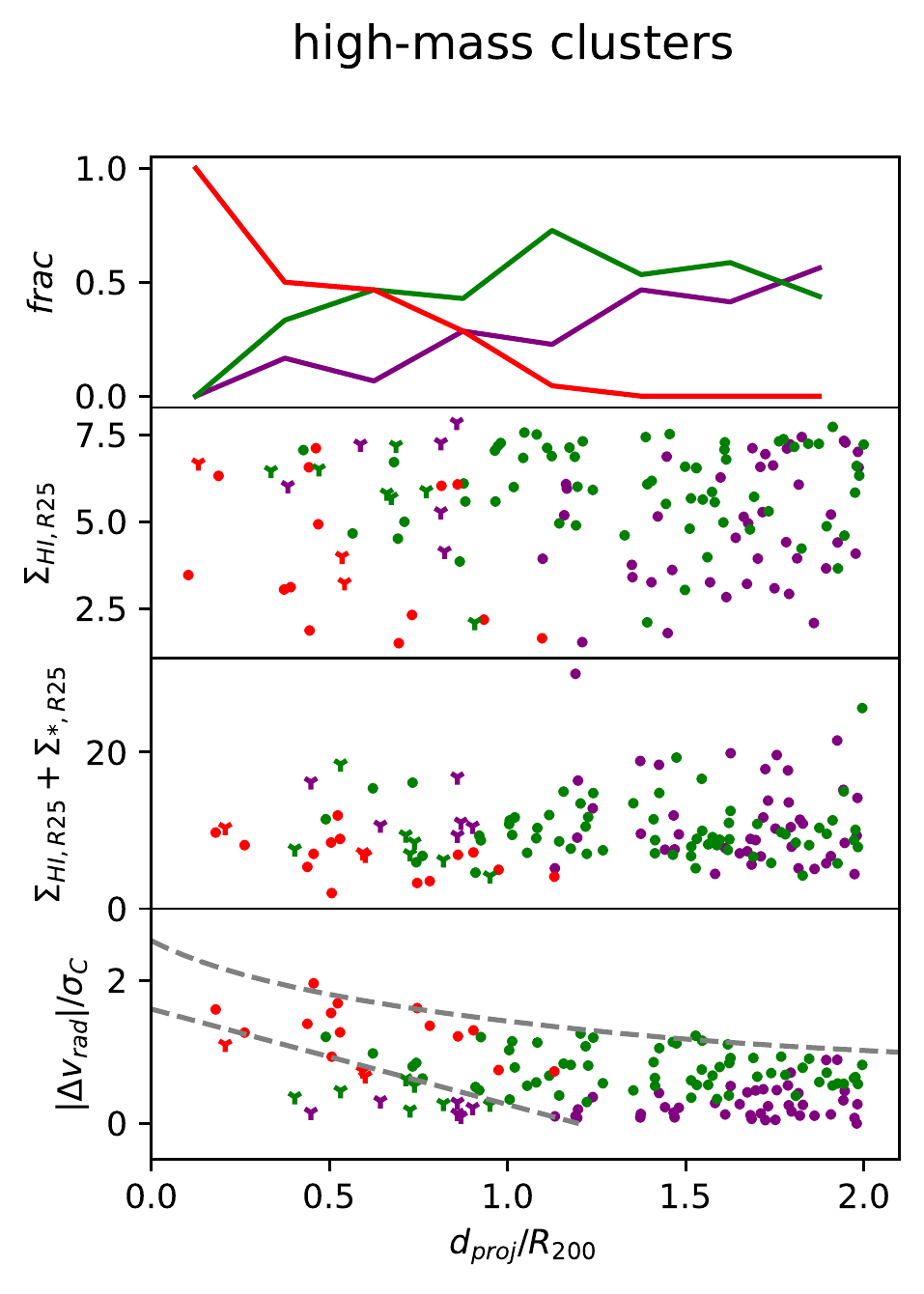}
\includegraphics[width=8.5cm]{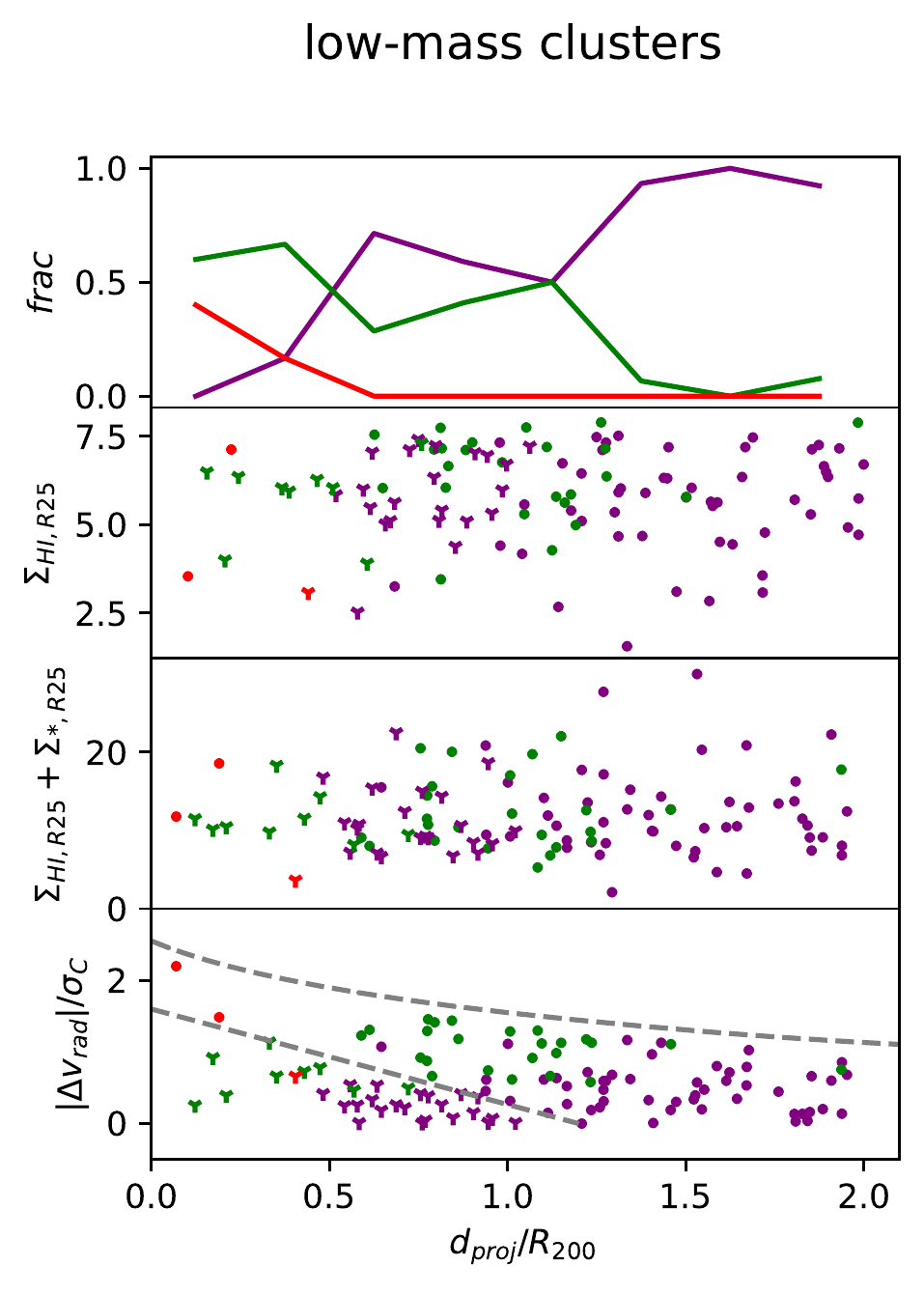}
\caption{ Distribution of galactic properties as a function of cluster centric radius. Galaxies have been identified as weak (green), strong (red) and no-RPS (purple) types, and further classified as within (tri-down markers) and beyond (circles) the virialized regions of clusters. From top to bottom, we plot the fraction of galaxies in each RPS type, the $\hi$ surface density at $R_{25}$ ($\Sigma_{{\rm HI}, R25}$, in unit of $M_{\odot}~pc^{-2}$), the sum of $\hi$ and stellar mass surface density at $R_{25}$ ($\Sigma_{{\rm HI}, R25}+\Sigma_{*, R25}$, in unit of $M_{\odot}~pc^{-2}$), and the projected PSD. In the projected PSDs, $v_{esc}$ assuming an NFW potential and the border of the virialized region \citep{Mahajan11, Jaffe15} are plotted as dashed lines. }
\label{fig:PSD}
\end{figure*}

In the bottom row of Fig.~\ref{fig:PSD}, the different RPS types separate well on the projected PSD. We further divide them into infalling and virialized types, and summarize their numbers in Tab.~\ref{tab:num_galaxies}. 
In Tab.~\ref{tab:num_galaxies}, the strong-RPS galaxies are not divided according to their infall status, because they are found continuously across the projected PSD. Most of the strong-RPS galaxies are found in the high-mass clusters, and in the infalling region. They dominate the $\hi$-rich galaxy population of high-mass clusters when $d_{proj}<0.5R_{200}$. 
The weak-RPS galaxies are also found mainly in the infall region, but extend further in high-mass clusters than in low-mass clusters.
They are rarely found beyond $d_{proj}/R_{200}\sim 1.2$ in low-mass clusters, but are common ($\sim$40\% among the three RPS types, top row of Fig.~\ref{fig:PSD}) at $d_{proj}/R_{200}\sim 2$ in high-mass clusters.
The no-RPS galaxies concentrate in the ``back-splash'' region ($d_{proj}>R_{200}$ and $\Delta v_{rad}/\sigma_C<1$) in high-mass clusters. They are dominantly found in two regions of the low-mass clusters:  the outskirts where $d_{proj}>1.2 R_{200}$, and the virialized region where $d_{proj}>0.5 R_{200}$. 

In other panels of Fig.~\ref{fig:PSD}, we can see that the strong-RPS galaxies tend to have low $\Sigma_{*,R25}+\Sigma_{\rm HI,R25}$ and low $\Sigma_{\rm HI,R25}$, hence low anchor forces at $R_{25}$. The weak and no-RPS galaxies are less different from each other in these properties.

 In summary, the RPS strength strongly depends on the cluster mass and the projected PSD position. Such strong dependence on PSD positions were noticed before \citep{Jaffe18, Yun19}.
 Although the RPS strength is determined by both the ram pressure and the anchor force, the former (reflecting the external environments) seems to play a dominant role in regulating the evolution of $\hi$-rich galaxies. The regulating role of anchor forces (reflecting the internal properties) rises only when the RPS is strong, probably after an earlier stage of weak RPS, which has reduced $\Sigma_{\rm HI, R25}$ and thus the anchor forces.
   
 \begin{table*}
\centering
{
\begin{tabular}{c|ccc|cc}
Cluster type & \multicolumn{5}{c}{Galaxy type} \\
 & \multicolumn{3}{c}{ Infalling }    & \multicolumn{2}{c}{Virialized} \\

& no-RPS & Weak-RPS &  Strong-RPS    & Weak-RPS & no-RPS \\
(1) & (2) & (3) & (4) & (5) & (6)   \\
\hline
High-mass&  42 & 63  & 17 & 7 & 6 \\
Low-mass&   51 & 24  & 3  &9& 22 \\

\end{tabular}
}
\caption{ {\bf Galaxy numbers in different types.} Column (1): type of clusters where the galaxies are found. Column (2)-(7): galaxy type. Column (2)-(4): Numbers of weak, strong and no-RPS galaxies within the virialized regions. Column (5)-(7): numbers of weak, strong and no-RPS galaxies beyond the virialized regions. }
\label{tab:num_galaxies}
\end{table*}

\subsection{HI mass and central SFR compared to control galaxies}
As the reservoir of material for star formation, the $\hi$ richness is strongly correlated with the star-forming status of field galaxies, and determines the future potential of forming stars. A lower $M_{\rm HI}$ than control galaxies which have similar integral SFR would be the signature of recent, violent removal of the $\hi$ gas, and predictive of a drop in SFR in the near future. 

Past studies \citep{Kauffmann03, Li15} showed that some spectral indices are good tracers of the recent star forming activities. Higher specific SFRs are typically associated with lower $D_{4000}$ and higher $EW(H\alpha)$. We remind that the control galaxies are matched in the {\it global} SFR to the main sample, and the SDSS spectral indices are measured for the galactic {\it central} regions. So if the main sample galaxies have lower central $D_{4000}$ and higher central $EW(H\alpha)$ than the control galaxies, it can be interpreted as these galaxies having higher central SFR, but lower SFR in the outer regions when compared to the control galaxies. Such an outside-in star formation cessation is often linked to gas stripping \citep{Koopmann04a, Koopmann04b, Boselli06, Boselli06a, Fossati13, Fabello12}.
On the other hand, if strong central starbursts happened in the past 10 Myr, galaxies are expected to have higher central $EW(H\alpha)$ for their central $D_{4000}$ \citep{Li15}. 

\begin{figure} 
\includegraphics[width=8.5cm]{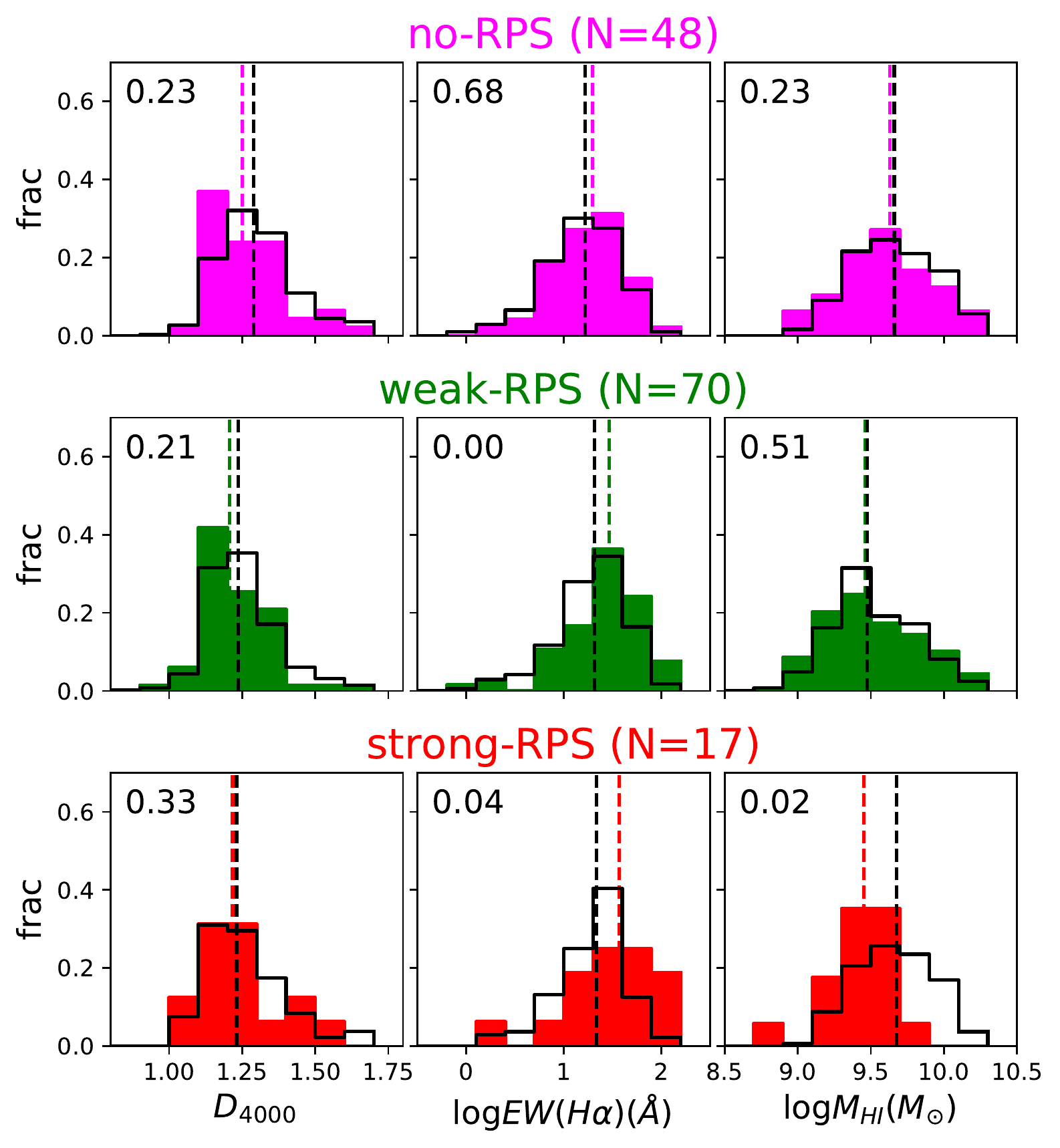}
\caption{ Comparison of star formation status and $\hi$ masses with control galaxies in high-mass clusters.  From the left to the right, the histograms of central $D_{4000}$, central $EW(H\alpha)$ and $\log M_{\rm HI}$ are plotted in each column. From the top to the bottom, we show histograms for the No- (magenta), weak- (green), and strong- (red) galaxies, respectively. The corresponding control sample for each sub-sample of cluster galaxies is plotted as black histograms, and the K-S test probability for the comparison between cluster and control galaxies are denoted in each panel. The median values are marked by the dashed lines. }
\label{fig:hist_Hmass}
\end{figure}

We compare the $M_{\rm HI}$ and central SFR of main and control samples in Fig.~\ref{fig:hist_Hmass} and \ref{fig:hist_Lmass}. 
We focus on the extent of differences each of the RPS sub-samples (strong, weak and no) shows with respect to their control galaxies. However, we will refrain from interpreting the apparent differences between the three sub-samples, because $M_{\rm HI}$ and SFR depends on several additional parameters (e.g. $M_*$, disk-bulge structure) which are not matched between the sub-samples.

Strong-RPS galaxies are shown in the third row of Fig.~\ref{fig:hist_Hmass}, which are mostly observed in the high-mass clusters. They tend to have on average higher central $EW(H\alpha)$ but similar central $D_{4000}$, indicative of enhanced central starbursts compared to their control galaxies.
They also have lower $M_{\rm HI}$ than the control galaxies. 

Three strong-RPS galaxies in the high-mass clusters have been excluded from this analysis because not enough control galaxies could be found (atlas in Fig.~\ref{fig:S_RPS_nocontrol} in the appendix). Two of these galaxies have significantly higher $SFR$ than all other galaxies with similar $z$, $M_*$ and $\mu_*$ in the control galaxy pool. The abnormalities in SFR are consistent with the enhanced central $H\alpha$ emission in the strong-RPS galaxies which have control galaxies.
The remaining galaxy without a control galaxy is an elliptical galaxy from the SDSS image. It has low integral $SFR\sim10^{-1.13}~M_{\odot}yr^{-1}$, but significant $M_{\rm HI} \sim 10^{9.3}~M_{\odot}$, which is 0.46 dex higher than expected for its $M_* \sim 10^{10.4} M_{\odot}$ and SFR \citep{Saintonge16}. Like in other $\hi$-rich early-type galaxies, its $\hi$ may have been obtained through mergers and maintained in the form of an extended disk, which does not easily flow to the galaxy center to fuel the star formation there \citep{Serra12}.

\begin{figure} 
\includegraphics[width=8.5cm]{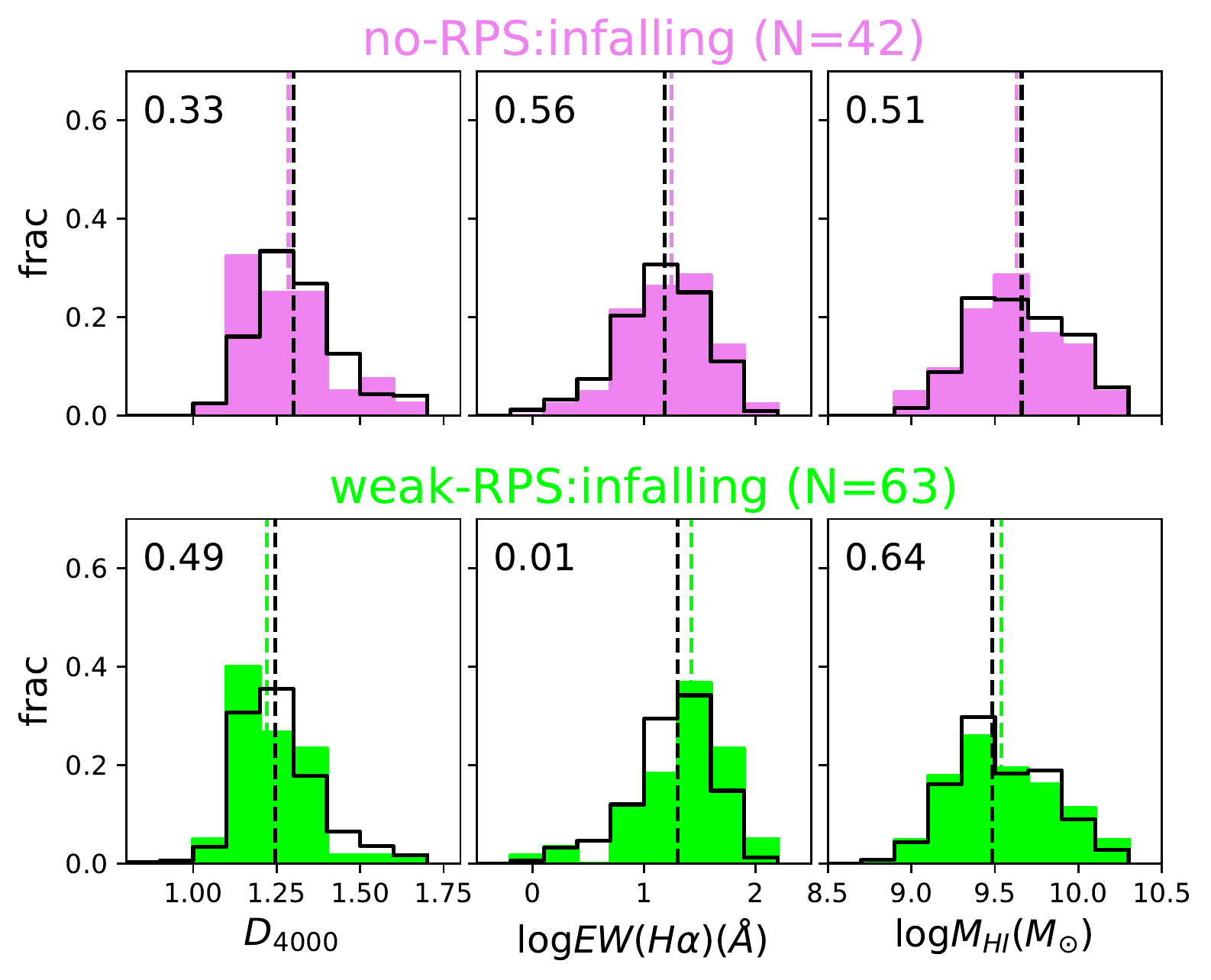}
\caption{ Distribution of star formation status and $\hi$ masses with control galaxies in high-mass clusters. Similar to Fig.~\ref{fig:hist_Hmass}, but only for infalling galaxies. We do not present figures for the virialized, weak- or no-RPS types, because of small sample sizes.  }
\label{fig:hist_divPSD_Hmass}
\end{figure}

\begin{figure} 
\includegraphics[width=8.5cm]{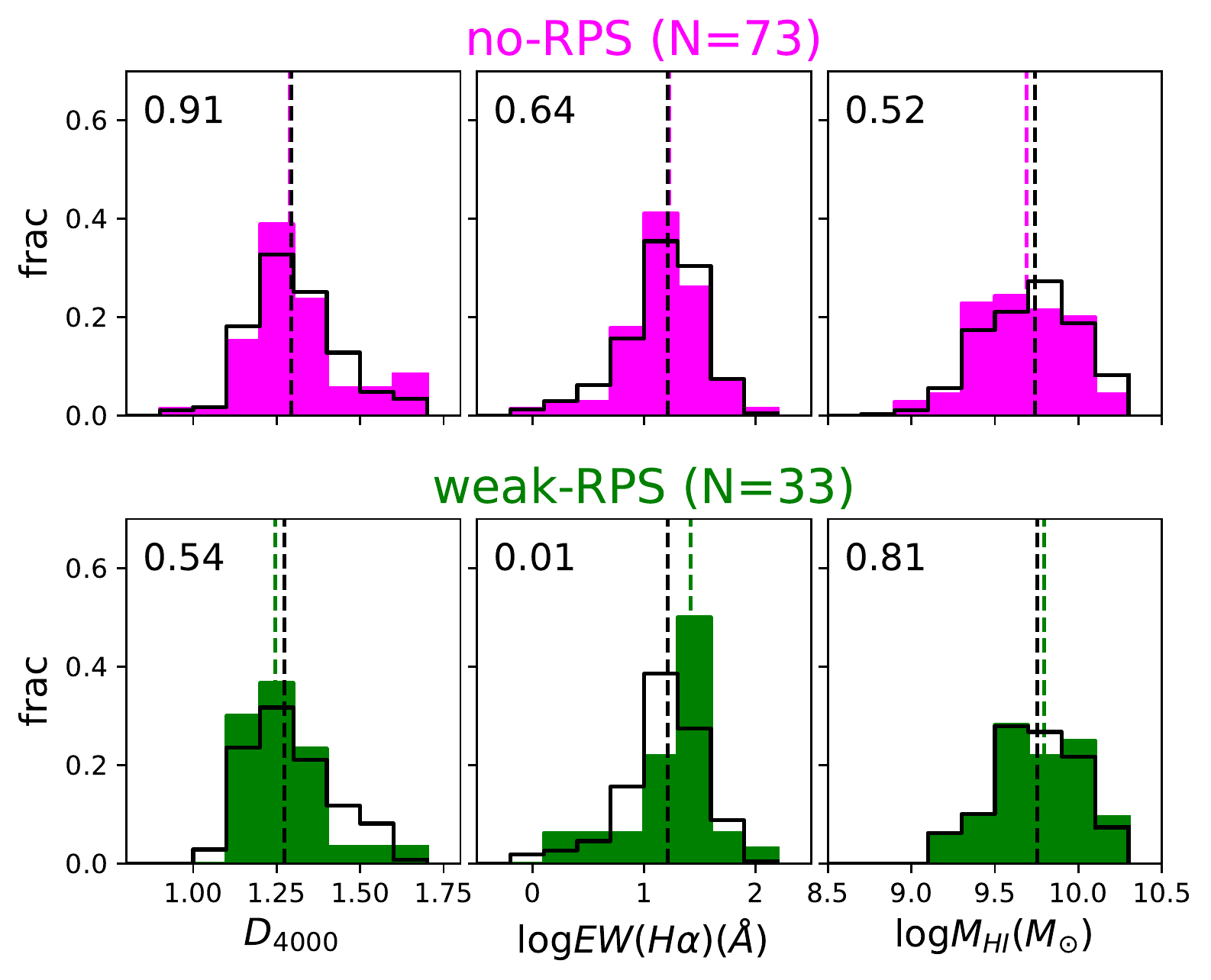}
\caption{ Distribution of star formation status and $\hi$ masses with control galaxies in low-mass clusters. Similar to Fig.~\ref{fig:hist_Hmass}, but for galaxies in low-mass clusters. We do not present figures for the strong-RPS galaxies because of small sample size.  }
\label{fig:hist_Lmass}
\end{figure}

\begin{figure} 
\includegraphics[width=8.5cm]{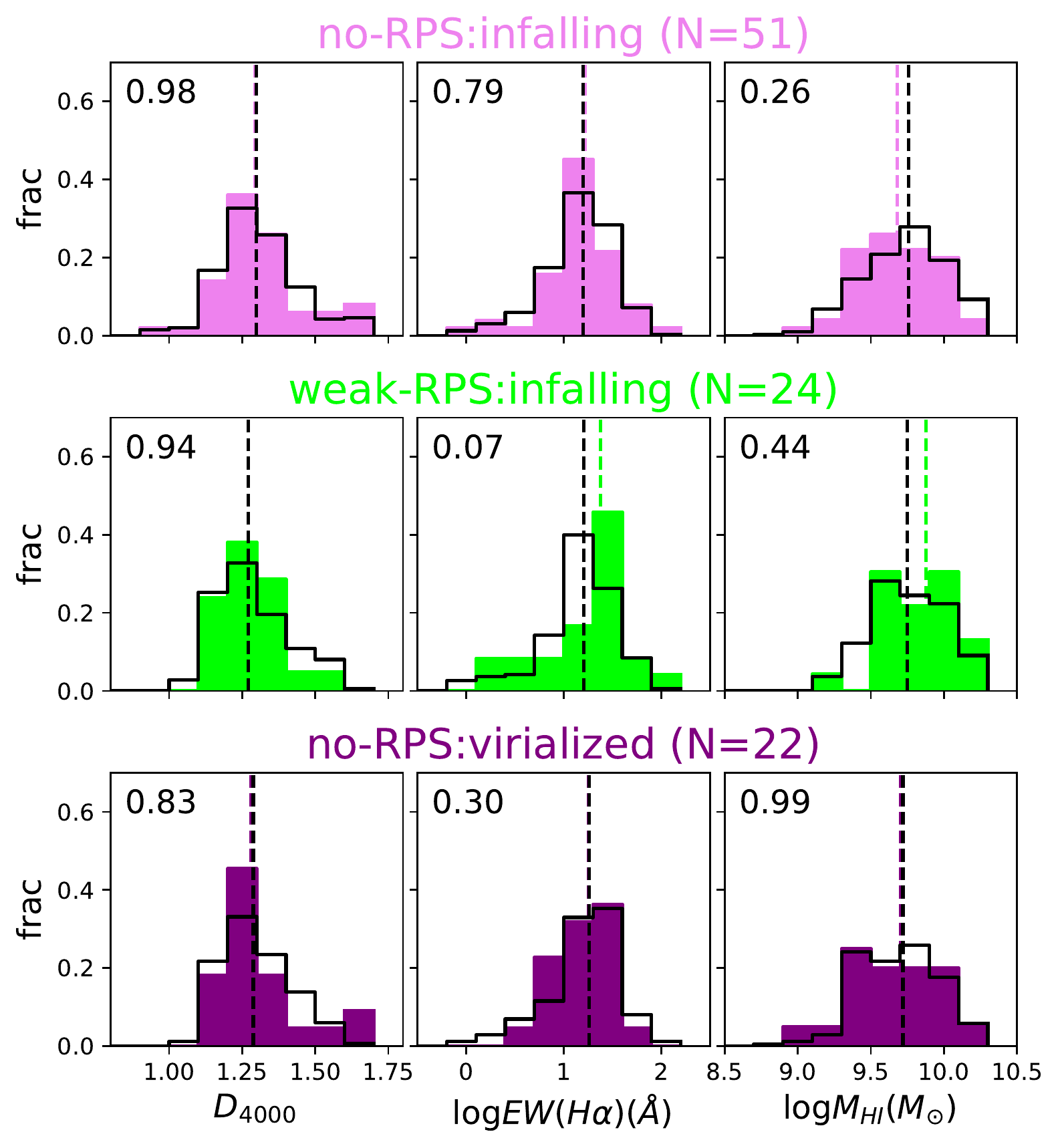}
\caption{ Distribution of star formation status and $\hi$ masses with control galaxies in low-mass clusters. Similar to Fig.~\ref{fig:hist_Lmass}, but no-RPS galaxies are further classified into infalling and virialized types, and we also present the infalliing, low-RPS galaxies. We do not present figures for the virialized, low-RPS galaxies because of small sample size. }
\label{fig:hist_divPSD_Lmass}
\end{figure}

The weak-RPS galaxies in the high-mass clusters have on average slightly higher central $EW(H\alpha)$ than the control samples (second row of Fig.~\ref{fig:hist_Hmass}). We find similar results when only selecting the infalling galaxies (second row of Fig.~\ref{fig:hist_divPSD_Hmass}), but do not have a large enough sample to conclude anything about the virialized weak-RPS galaxies. Similar results are found in the low-mass clusters (second row of Fig.~\ref{fig:hist_Lmass}, and second row of Fig.~\ref{fig:hist_divPSD_Lmass} ). None of the weak-RPS subsets show significantly different $M_{\rm HI}$ distribution compared to the control galaxies.

The no-RPS galaxies do not significantly differ from their control galaxies in the distributions of central $EW(H\alpha)$, central $D_{4000}$, or $M_{\rm HI}$. This result holds when selecting either virialized or infalling galaxies, in both low-mass and high-mass clusters (Fig.~\ref{fig:hist_Lmass}, Fig.~\ref{fig:hist_Hmass} and Fig.~\ref{fig:hist_divPSD_Lmass}). 


\section{Discussion}
\label{sec:discussion}
Combining HI data of galaxies with extended X-ray data of clusters was conducted for individual clusters (e.g. \citealt{Jaffe15}) before, but this paper for the first time works on a relatively complete overlap between the largest $\hi$ blind survey ALFALFA and the largest X-ray blind survey ROSAT. The galaxies were selected from two complete catalogs of clusters with extended X-ray emissions; particularly the RXGCC sample is lately built with a noval algorithm to search for faint clusters from the whole ROSAT dataset. The extended X-ray fluxes ensures $M_{200}$, $R_{200}$ and $n_{ICM}$ to be derived in relatively accurate ways. The results presented in this study can thus be compared to simulations in a relatively convenient way in the future, by producing mock catalogs with similar survey parameters as ALFALFA and ROSAT. 

We point out that, the major goal of this study is neither characterizing galactic features under RPS, nor providing a census of galaxies under strong RPS, for the sample is strongly biased against $\hi$ deficient galaxies. Instead, we examine whether the observed $\hi$ and SFR properties are consistent with expectations when galaxies are under RPS, as a test of our classification method (Sec. 5.2); we discuss the role that weak RPS might play in cluster galaxy evolution, based on the PSD distribution and frequency of weak-RPS galaxies in our sample (Sec. 5.3).

\subsection{Past studies on HI and SFR properties of galaxies in the Coma and A1367 clusters}
We note that a considerable fraction of the main sample galaxies come from the Coma and A1367 clusters. The $\hi$ and SFR properties of galaxies in these two clusters have been extensively studied before.

Studies based on wide-field, blind $\hi$ surveys found that, the distribution of HI detected galaxies is much less concentrated on the cluster centers than optically detected galaxies \citep{Cortese08}, and galaxies are more $\hi$-deficient when the local densities are higher \citep{Gavazzi13}.
 Studies based on ultraviolet, infrared and H$\alpha$ images consistently found the outside-in suppression of SFR at high local densities \citep{Cybulski14, Gavazzi13}. 
  These statistical results support a picture where ram pressure plays an effective role in removing $\hi$ from galaxies. 

Interferometric $\hi$ images further confirmed the on-going RPS for a number of galaxies. The $\hi$ disks of observed galaxies in the Coma and A1367 clusters often display lopisided morphologies, displaced center from the optical counterparts, and$\slash$or smaller extension than the optical disks \citep{BravoAlfaro00, BravoAlfaro01,Scott10, Scott18}. The ubiquitous RPS in the Coma cluster was also confirmed from observing the warm ionized gas with deep H$\alpha$ images  \citep{Gavazzi18, Yagi17, Yagi10}.

\subsection{ Relating the observed trends in this study to RPS}
 We find that, only the strong-RPS galaxies show evidence for a significant reduction in $M_{\rm HI}$ compared to the control galaxies; strong- and weak-RPS galaxies show higher central $EW(H\alpha)$  in the galactic center than the control galaxies; no-RPS galaxies are no different from the control galaxies in either $M_{\rm HI}$ or the central $EW(H\alpha)$. We discuss possible physical mechanisms relating RPS to these observed differences. 

\subsubsection{Differences in $M_{\rm HI}$}
The strong-RPS galaxies have on average lower $M_{\rm HI}$ than the control galaxies, suggesting a fast removal of $\hi$. 
$\hi$ does not directly form stars, but fuels star formation as part of the circle of gas accretion, gas inflow, star formation and outflow \citep{Krumholz18, Wang20}. So at a given $M_*$, SFR is adjusted to the available $\hi$ on a timescale longer than the free fall time of molecular gas \citep{Krumholz12}, but shorter than the $\hi$ depletion time \citep{Saintonge17}. Under strong RPS, the gas removal can be much quicker than the capability for SFR to be adjusted to $M_{\rm HI}$. For extreme cases of galaxies in the stripping region of the projected PSD, gas removal has a timescale of a few $10^7$ yr \citep{Abadi99}.  
 Our result is consistent with previous findings that the detection rate and mass fraction of $\hi$ drops much more quickly than the specific SFR near the cluster centers \citep{Fabello12, Jaffe15}. Cross-matching with nearby galaxies which are known to display RPS gas tails, we confirm that at least 4 out of the 17 strong-RPS galaxies are indeed among those with tails (see Sec.~\ref{sec:appendix_LV} of Appendix). This confirmation rate should be viewed as a lower limit, for not all our clusters have been searched for RPS features in the morphology before. Gas removal in low- and no-RPS galaxies seems to be much slower. One direct reason is likely that a smaller radial range of $\hi$ is affected in low-RPS galaxies than in high-RPS galaxies. Additionally, there is a time lag between gas being stripped off the disk plane and reaching the escape velocity of the galaxy, which is longer when the ram pressure is weaker \citep{Roediger07}. In addition to RPS, tidal stripping or harassment may also contribute to reducing $M_{\rm HI}$ in the weak- and no-RPS galaxies because their velocities are lower than those of the strong-RPS galaxies. Yet the combined efficiency of removing $\hi$ is not as high as in the strong-RPS galaxies. 

\subsubsection{Differences in the central SFR}
The strong-RPS and low-RPS galaxies show higher central $EW(H\alpha)$  in the center when compared to their control galaxies, consistent with the consequence of RPS. Higher values of central $EW(H\alpha)$ compared to the control sample could indicate either recently enhanced central SFRs, or suppressed SFRs in the outer disks. If the strong-RPS or low-RPS galaxies had on average lower central $D_{4000}$ values than their control galaxies, then it would be strong evidence for suppressed SFR in the outer region. 
But we observed no significant difference in central $D_{4000}$ between the low-RPS (strong-RPS) galaxies and their control galaxies. Considering the fact that $D_{4000}$ may not be as sensitive as $EW(H\alpha)$ to small changes in the sSFR, we discuss both possibilities regarding whether SFR is suppressed in the outer disks. 

If the SFR is indeed suppressed in the outer region of strong- and low-RPS galaxies, then it strongly supports a scenario of outside-in quenching as a result of outside-in gas stripping. Such a stripping scenario is consistent with the nature of RPS, because the anchor force is weaker at larger galactic radius. RPS could then perfectly explain the central $EW(H\alpha)$ enhancements in strong- and weak-RPS galaxies compared to no-RPS galaxies. 

It is also plausible that the strong- and weak-RPS galaxies have enhanced central SFR instead of (or in addition to) suppressed outer SFR.
Theoretical studies predicted the enhancement of SFR when pressure from the ICM or shocks at the ICM-disk interface compress the cold gas, before the gas is severely stripped \citep{RamosMartinez18, Safarzadeh17, Steinhauser16}. In several simulations the process is accompanied by significant gas inflows, generated directly by oblique shocks \citep{RamosMartinez18}, or loss of angular momentum in interaction with the ICM \citep{Tonnesen09}, which results in enhanced central SFR \citep{Bekki14, Tonnesen12, Kronberger08}. Observational studies also found enhanced SFR prevalent in galaxies undergoing ram pressure \citep{Roberts20,Vulcani18, Jaffe16}. But the preferred location within galaxies for SFR to be enhanced is debating, which can be in the ICM-disk interface \citep{Ramatsoku19, Lee17, Ebeling14}, in the center \citep{Mok17}, and at all galactic radius \citep{Vulcani20}. 
 The enhanced central SFR in strong-RPS galaxies is thus not against the literature findings.
 
 But the mechanism of enhancing central SFR might be more complex in the weak-RPS galaxies. As the low-RPS galaxies tend to be in the relatively outer region of the cluster, tidal interactions with both the cluster \citep{Byrd90} and surrounding galaxies \citep{Mihos92} might also play a role by driving gas inflows. However, interestingly, in low-mass clusters, the virialized, no-RPS galaxies do not show enhanced central SFR, although they are in a similar $d_{proj}$ range and thus similar cluster gravities and local densities as the infalling, low-RPS galaxies. These virialized, no-RPS galaxies should even suffer from more effective tidal interaction with the surrounding galaxies, due to their lower velocities than the infalling, low-RPS galaxies. Yet they do not show as much enhanced central SFR as the infalling, low-RPS galaxies. It is possible that galactic tidal effects even in the outer region of these massive clusters are generally weak \citep{Boselli06}, and take the form of harassment \citep{Moore96, Moore98} instead of interactions, which heat the disks but do not efficiently drive gas inflows. Meanwhile, tidal interaction with the cluster may not be so efficient at these relatively large distances, and indeed we find that all of the infalling, low-RPS galaxies have a cluster perturbation strength $(M_{200}/M_*)((\pi/4)*(R_{25}/d_{proj}))^3$ lower than the critical value of 0.1
\footnote{This critical value of 0.1 assumes that the galaxy preserves its dark matter halo \citep{Byrd90}, which is reasonable as 90\% of the infalling, weak-RPS galaxies have tidal radius $r_{tid}\sim0.5W_{50}/\sigma_C>8R_{25}$ \citep{Merritt84} in both low and high-mass clusters. The tidal radius has been defined as the galactic radius beyond which material is effectively removed by tidal effects. For reference, if we assume that no dark matter remains, the critical value of cluster tidal perturbation strength drops to 0.006 \citep{Byrd90}, and 70\% (38\%) of infalling, weak-RPS galaxies in high-mass (low-mass) clusters are perturbed according to this criteria. } 
for triggering nuclear activities \citep{Byrd90}. It implies that ram pressure may be the mechanism that enhanced the central SFR in the weak-RPS galaxies. 

\subsubsection{Feasibility of the RPS strength parameter}
As discussed above, cluster galaxies under stronger RPS exhibit more significant difference in $M_{\rm HI}$ and central SFR from their control galaxies. 
It is worth noting that the three RPS types occupy different regions of the projected PSD in high-mass and low-mass clusters. The difference is expected because $\sigma_C$ of the high-mass clusters is higher, leading to higher $\Delta v_{rad}$ at a given projected PSD position than in the low-mass clusters. 
But, the low-mass and high-mass clusters show consistent results when comparing the different RPS types to control galaxies. It implies that the RPS strength parameter has captured a fundamental property of the environmental processing. Although the RPS strength parameter combined several observables of external and internal properties, the way these observables are combined is not arbitrary but motivated by the RPS theory. Although the way of inputting the observables to the calculation of PRS strength has uncertainties, it seems that the physical effect of RPS is strong enough in the massive clusters to override many of the uncertainties. 

Thus, we conclude that our classification of galaxies into the three RPS types is statistically successful. 

The way we define the three populations can be effectively used to combine galaxies from different clusters. In statistical analysis, cluster galaxies were often binned into sub-sample by more directly observed parameters like $d_{proj}/r_{200}$; here we have introduced $P/F_{R25}$ as a new parameter to bin sub-samples. This new parameter can be conveniently derived in cosmological simulations, and helps separate RPS from other processes$\slash$parameters that influence galaxy evolution.

\subsection{ Radial extension of RPS in the massive clusters}
Early statistical studies on environmental effects in groups$\slash$clusters focused on the cluster centric trend of galaxy properties (or similarly, galaxy properties as a function of local densities). Those studies found that galaxies on averaged become more $\hi$-deficient and passive toward the cluster center \citep{Brown17, Odekon16, Yoon15, vonderLinden10, Gavazzi10, Gavazzi06, Weinmann06}, implying accelerated galaxy evolution in clusters. Such a trend is found to be steeper in high-mass clusters than in low-mass clusters \citep{Brown17, Yoon15, Hess13, Woo13}, more significant for low-mass galaxies than for high-mass galaxies \citep{Woo17, Zhang13, Woo13, Wetzel13}, consistent with the way that RPS is predicted to work.
Later, it was found that galaxy properties also vary as function of radial velocity offsets from the cluster center at a given projected distance \citep{Nascimento19, Bayliss17, Barsanti16, Mahajan11, Pimbblet06}. 
With the aid of cosmological simulations, it becomes clear that positions on the PSD are associated with galaxies at different infall stages, thus show correlation with the averaged galactic properties \citep{Haines15, Boselli14a, Gill05}.

The PSD becomes an excellent tool to study RPS, not only because it can identify infalling galaxies which may have more gas to be stripped \citep{Rhee17, Haines15, Oman13}, but also because its two axes almost fully determine the ram pressure for a given cluster \citep{Gunn72}. In observations, only the projected PSD is available, but has been proven to be statistically powerful in linking infall stages, stripping events, gas-richness and star-forming status of galaxies \citep{Yoon17,  Oman16, Boselli14a, Muzzin14}.

The pioneer studies utilizing the projected PSD to study RPS of $\hi$ in galaxies found remarkable consistency between observed and predicted $\hi$-richness. These studies typically assume exponentially radial distributions for both the $\hi$ and stars, and the scale-length of an $\hi$ disc is set to be a fixed factor of that of the stellar disk. Then, basing on the equation of \citet{Gunn72} and setting the relative velocity to be $\sigma_C$, a limiting ``stripping region'' could be defined in the projected PSD where the ram pressure becomes stronger than the anchor force at all galactic centric radii, and thus the gas in galaxies is expected to be significantly removed within this region. The detection rates of galaxies in blind $\hi$ surveys abruptly drop, and the fractions of red galaxies significantly increase after passing that limit \citep{Jaffe15, Jaffe16, Yoon17, Jaffe18}. These results lent strong support to RPS driving galaxy evolution in massive clusters, in a more direct way than previously using cluster centric radial trends. 

\begin{figure*} 
\centering
\includegraphics[width=16cm]{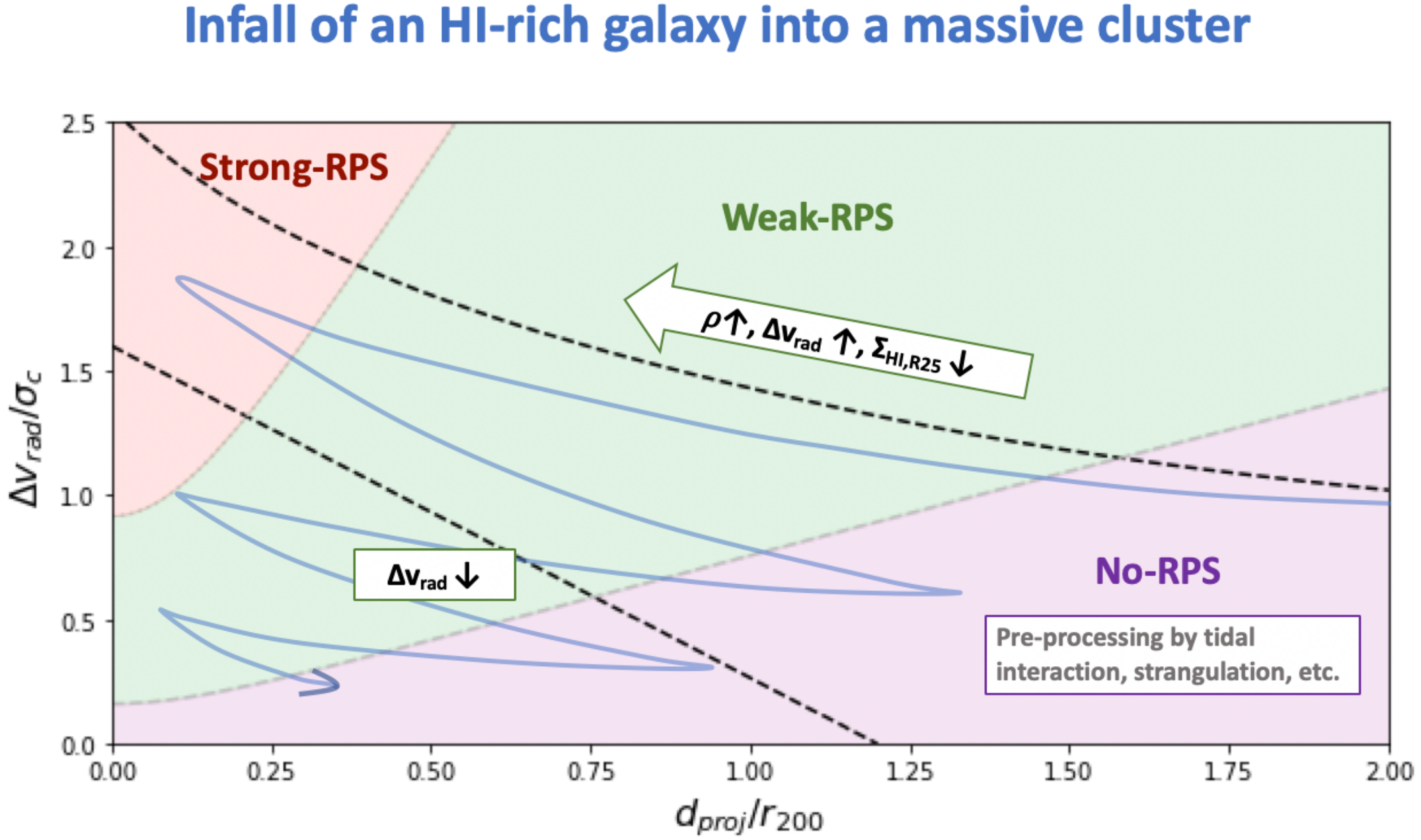}
\caption{Toy scheme of an $\hi$-rich galaxy passing different RPS regions while traveling through a massive cluster. Like in Fig.~\ref{fig:PSD}, the black, dashed curves mark the escape velocity and the virialized region. The arrowed blue curve is the trajectory of the galaxy starting from $d_{proj}\sim 2 R_{200}$, shrinking due to dynamic friction until getting virialized. The purple, green and red regions are divided by curves of equivalent ram pressure, and approximately correspond to the no, weak and strong-RPS regions, because RPS strengths are largely determined by the relative velocity of the galaxy and the density of the ICM. But the gradually reduced anchor force due to the drop of $\Sigma_{{\rm HI}, R25}$ also enhances the RPS strength, particularly when the galaxy enters the strong-RPS region. }
\label{fig:scheme}
\end{figure*}

Our work is built upon these previous analyses with two new components added to the method. The first new component is that we use a more realistic radial distribution of $\hi$ when estimating the anchor forces. Compared to the exponential model often assumed in the previous studies, a real $\hi$ disk tends to have a flattened surface density distribution, and sometimes a central hole in the inner region. Thus for the same $M_{\rm HI}$ and scale-length, a real disc tends to have more mass and hence higher surface densities in the outer region, resulting in higher anchor forces. Additionally, statistical analysis based on $\hi$ images found that the $\hi$ scale-length$\sim0.2R_{\rm HI}$ \citep{Wang14, Wang20}, which strongly depends on $M_{\rm HI}$ but not on the optical scale-length. The ratio between the $\hi$ and $r-$band scale-lengths of our main sample galaxies ranges from 0.7-1.9 in both high-mass and low-mass clusters. Assuming a fixed ratio of the scale-lengths as in the previous studies will thus introduce additional uncertainties in $\hi$ surface densities at a given galactic radius. The second new component of the method is that we focus on an earlier phase of stripping, when ram pressure is just enough to strip the $\hi$ at $R_{25}$ and $R_{\rm HI}$. The selection of galaxies with $R_{\rm HI}>R_{25}$ also ensures that most galaxies have not entered the classical ``stripping region'' yet. We thus included the self-gravity of $\hi$ when calculating the anchor forces, which is usually ignored when discussing stripping of the inner disks where stars dominate the gravity. We also used $\Delta v_{rad}$ instead of $\sigma_C$ when calculating the ram pressure, to better reflect the fact that during infall galaxies are accelerated while approaching the pericentre. 
 A few interesting features show up with this new scheme of classifying galaxies into strong-, weak- and no-RPS populations. We summarize the scenario in Fig.~\ref{fig:scheme} and discuss a few key points below.

First, strong-, weak- and no-RPS galaxies overlap significantly in $d_{proj}$. Thus the scatter in the previously quantified cluster centric radial trends of gas richness \citep{Brown17, Odekon16, Hess13} can be explained at least partly by this feature. As already mentioned, at a given $M_{200}$, the projected PSD position traces the ram pressure of different levels much more closely than $d_{proj}$. The higher incidence of RPS for the infalling galaxies (with respect to the virialized galaxies) was also noticed in previous studies \citep{Jaffe18, Yun19}.
Whether the RPS is effective further depends on the anchor force determined by the gas and stellar surface densities in the galaxy. The RPS strength (the ram pressure over the anchor force) as one parameter puts together these factors in a physically motivated way. Despite the projection effects and other uncertainties discussed in Sec~\ref{sec:analysis}, their statistical correctness is supported by the distinct $\hi$ and SFR properties of the three RPS populations. 

More importantly, the strong- and particularly weak-RPS galaxies make a significant fraction ($\sim$ half) of the $\hi$-rich sample with $d_{proj}$ extending out to at least $R_{200}$, and even to $2R_{200}$ in high-mass clusters. Evidence for widely distributed RPS out to at least $R_{200}$ was reported in the Coma Cluster \citep{Gavazzi18, Roberts20}. Jellyfish galaxies from the project GASP are also found out to $R_{200}$ \citep{Jaffe18}. However, limited by imaging efficiencies (and the possibility that weak-RPS only produces weak signatures), there was no observational census yet regarding the fraction of gas-rich galaxies undergoing RPS. The significance of RPS in the outskirts of clusters has therefore been rarely discussed in observations. 
The weak RPS may remove $\hi$ much more slowly than strong-RPS (which has produced observable reduction in $M_{\rm HI}$ at a given SFR), so their SFR can catch up with the change in $M_{\rm HI}$. But weak RPS is not necessarily much slower than the tidal stripping from the cluster which has a timescale similar to the crossing time of the cluster ($\sim$2 Gyrs, \citealt{Boselli06}). The SFR enhanced by the weak  ram pressure further consumes the gas.
 As the velocity is accelerated and the ICM density increases, the RPS will successively strengthen during the infall of a galaxy.  We may not need to wait until the galaxies reach the stripping region, before the cumulative effect from RPS has significantly affected galaxy evolution. As predicted by hydro-dynamic simulations, the combined effect of stripping and consumption due to ram pressure between a cluster centric distance of $R_{200}$ and $1/3R_{200}$ can account for one third of the total gas loss in a galaxy during the infall process between $R_{200}$ and the pericentre (\citealt{Steinhauser16}, their Fig.2 and 4). At this point, we are still unable to directly compare in observations the relative importance of RPS to tidal stripping of the cluster and other environmental effects (harassment, viscosity stripping, thermal evaporation, etc.), but we showed that for a significant fraction of gas-rich galaxies at and beyond $R_{200}$ of massive clusters, ram pressure is likely already causing $\hi$ loss. The high incidence of RPS in $\hi$-rich galaxies at $R_{200}$ is consistent with the prediction of the simulation IllustrisTNG-100 \citep{Yun19}. Environmental processing beyond $R_{200}$ is commonly termed pre-processing and attributed to effects in groups \citep{Bahe19, Bahe15, Bahe13}, but our results suggest that part of the ``pre-processing'' around the most massive clusters could actually be processed by the cluster itself through weak RPS.

\subsection{Uncertainties and future perspective}
\begin{figure} 
\includegraphics[width=8.5cm]{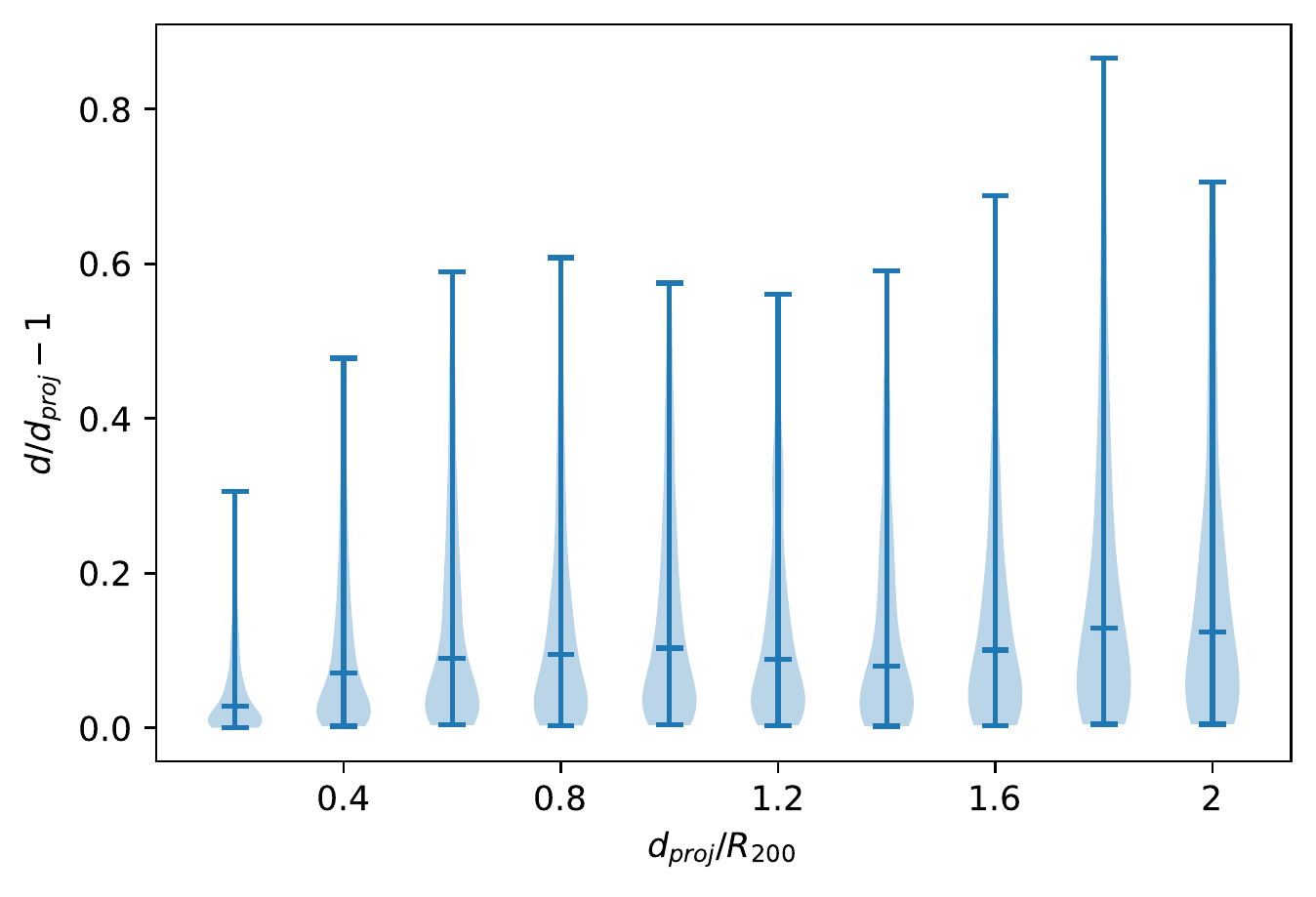}
\caption{ Violin plot of the difference between $d$ and $d_{proj}$ of galaxies in clusters selected from the TNG300 run in the IllustrisTNG project. Only galaxies with mass$>10^9 M_{\odot}$ are selected. The three bars of each violin represent the 10, 50 and 90 percentiles of the distribution. The distributions do not change much if we only select infalling galaxies, or exclude the backsplash galaxies. }
\label{fig:dproj}
\end{figure}

We warn readers again about the uncertainties related to the estimates of the RPS parameter. Among those discussed in the paper, the most obvious one is the projection effect. Luckily, as massive clusters strongly concentrate galaxies, a $d_{proj}$ is associated with a relatively narrow distribution $d$ with the median value slightly larger than but close to $d_{proj}$ (in contrast, in the field $d_{proj}$ and $d$ are much more different). Similarly, $\Delta v_{rad}$ is much closer to $\Delta v$ than they would be in looser environment. 
We roughly quantify the difference between $d_{proj}$ and $d$ by selecting all the 280 clusters with $M_{200}>10^{14}~M_{\odot}$ (having in total 23295 galaxies with mass$>10^9~M_{\odot}$) from the TNG300 run in the suite of IllustrisTNG cosmological simulations \citep{Nelson19, Springel18}. From Fig.~\ref{fig:dproj}, we can see that at all $d_{proj}$, the majority ($>80\%$) of the cluster galaxies have $d$ differ by less than 40\% from $d_{proj}$ (see also \citealt{Mahajan11}). This is one important reason why with projected distances, the estimated ram pressure still statistically select galaxies with SFR and $\hi$ properties consistent with the expected RPS.
We roughly assess the uncertainty of approximating $d$ with $d_{proj}$, by replacing $d_{proj}$ with $1.5d_{proj}$ when estimate $\rho$. Then, $\sim$45\% of the main sample galaxies are under weak RPS at $1.6 R_{200}$ in the high-mass clusters; only $25\%$ of the main sample galaxies, but nearly half of the infalling subset are under weak RPS at $R_{200}$ in the low-mass clusters. Because $\Delta v_{rad}$ is an under-estimate of $\Delta v$, the real ram pressure is likely stronger, and more galaxies may be under weak RPS at these distances than classified with $\Delta v_{rad}$. 
So our main result that weak-RPS affects a significant fraction of galaxies near and beyond $R_{200}$ is likely robust. 

We emphasize that application of the method should always be limited to statistical analysis, and future comparison with hydro-dynamic and semi-analytical simulations may help us quantify and correct for the limitations. 
Most previous comparisons between the observation and the simulation were in the form of comparing scaling relations and cluster centric radial distribution of observable parameters, which reflect the result of complex physical processes mixed together. 
Our RPS strength parameter provides an opportunity to (at least partly) separate RPS from other environmental and internal processes, in such comparisons. For example, recent $\Lambda$CDM semi-analytical models (SAM) found that RPS of cold gas is a necessary component in the model to reproduce the observed level of $\hi$ mass fractions in relatively high-mass satellite galaxies, but the $\hi$ mass fraction of low-mass satellite galaxies, and the offsets of $\hi$ related scaling relations between central and satellite galaxies are difficulty to reconcile \citep{Cora18, Stevens17, Luo16, Henriques15, Gonzalez-Perez14}. Directly comparing to the observed distribution of RPS strength, and to the observed $\hi$ property as a function of RPS strength may help identify whether and which part of the RPS recipe in SAMs need to be improved. 

 The sample used in this study is still relatively small, thus the differences between strong$\slash$weak-RPS and control galaxies are marginal in each individual figure of Fig.\ref{fig:hist_Hmass}-\ref{fig:hist_divPSD_Lmass}. The differences remain consistent in all these figures, adding strength to their statistical significance, but they need confirmation with larger samples in the future.

Finally, as in many other studies utilizing ALFALFA data \citep[e.g., ][]{Odekon16}, we are limited by the depth of $\hi$ data, which biased the sample against low-mass galaxies. We look forward to deeper and more $\hi$ detections with the up-coming CRAFTS \citep{Zhang19}, Apertif \citep{Verheijen09}, and WALLABY \citep{Koribalski20} surveys. $\hi$ images are usually used to search for RPS tails, but as the observability of tails depends on the angle of the line of sight from the direction of infall as well as the image resolution, our method can be used to select RPS candidates that do not show obvious tails. Such an application is particularly useful when considering that the majority (90\%) of galaxies going to be detected in those new $\hi$ survey will not be well resolved \citep{StaveleySmith15}. 

\section{Summary and conclusion}
\label{sec:conclusion}
So far as we know, this is the first statistical study on ALFALFA detected galaxies in more than ten clusters with well parametrized extended X-ray emissions (i.e. with resolved X-ray surface brightness radial profiles). The sample of clusters extends to $M_{200}\sim 4\times10^{13} M_{\odot}$ thanks to the new RXGCC catalog (\citealt{Xu18}, Xu et al. in prep) built with the state-of-art algorithms searching for faint and extend X-ray sources.
  
We described a promising method to parametrize the RPS strength in clusters, based on the theory of \citet{Gunn72} and improved upon the previous observational achievements. We compared the $M_{\rm HI}$ and central SFR of over 200 $\hi$-rich cluster galaxies to a control sample of field galaxies which are matched in the total SFR, $M_*$, stellar surface density and redshift. 

We showed that galaxies under stronger RPS also have faster $\hi$ removal and more enhanced central $SFR$ (compared to general galaxies with similar global $SFR$). The trend holds for both infalling and virialized galaxies, and in both low-mass and high-mass clusters, implying that the parameter successfully indicates the RPS strength. 
Because the weak-RPS is likely to extend beyond $R_{200}$ for a significant fraction of $\hi$-rich, infalling galaxies, processing and pre-processing may not have a clear border at $R_{200}$ for the most massive clusters. Because it works in a wide cluster centric radial range, weak RPS may play a significant role in galaxy evolution in massive clusters. 
Our parameter of RPS strength brings observables closer to theoretical models of RPS, which may help disentangle RPS from other environmental effects on galaxy evolution in future applications. 

\acknowledgments
We gratefully thank Martha Haynes for generously providing the ALFALFA $\hi$ spectrum. JW thank support from National Science Foundation of China (11721303).  

GALEX (Galaxy Evolution Explorer) is a NASA Small Explorer,
launched in April 2003, developed in cooperation with the
Centre National d'Etudes Spatiales of France and the Korean Ministry
of Science and Technology.

We thank the many members of the ALFALFA team who
have contributed to the acquisition and processing of the ALFALFA
dataset over the last many years. RG and MPH are supported by NSF
grant AST-0607007 and by a grant from the Brinson Foundation.

Funding for the SDSS and SDSS-II has been provided by
the Alfred P. Sloan Foundation, the Participating Institutions, the
National Science Foundation, the U.S. Department of Energy,
the National Aeronautics and Space Administration, the Japanese
Monbukagakusho, the Max Planck Society, and the Higher Education
Funding Council for England. The SDSS Web Site is
http://www.sdss.org/.

\bibliographystyle{apj}
\bibliography{HI_Xraycluster}

\appendix
\section{More information about the strong-RPS galaxies}
\label{sec:appendix_srps}
\subsection{The SDSS atlas}
\label{sec:appendix_atlas}
\begin{figure*} 
\centering
\includegraphics[width=15cm]{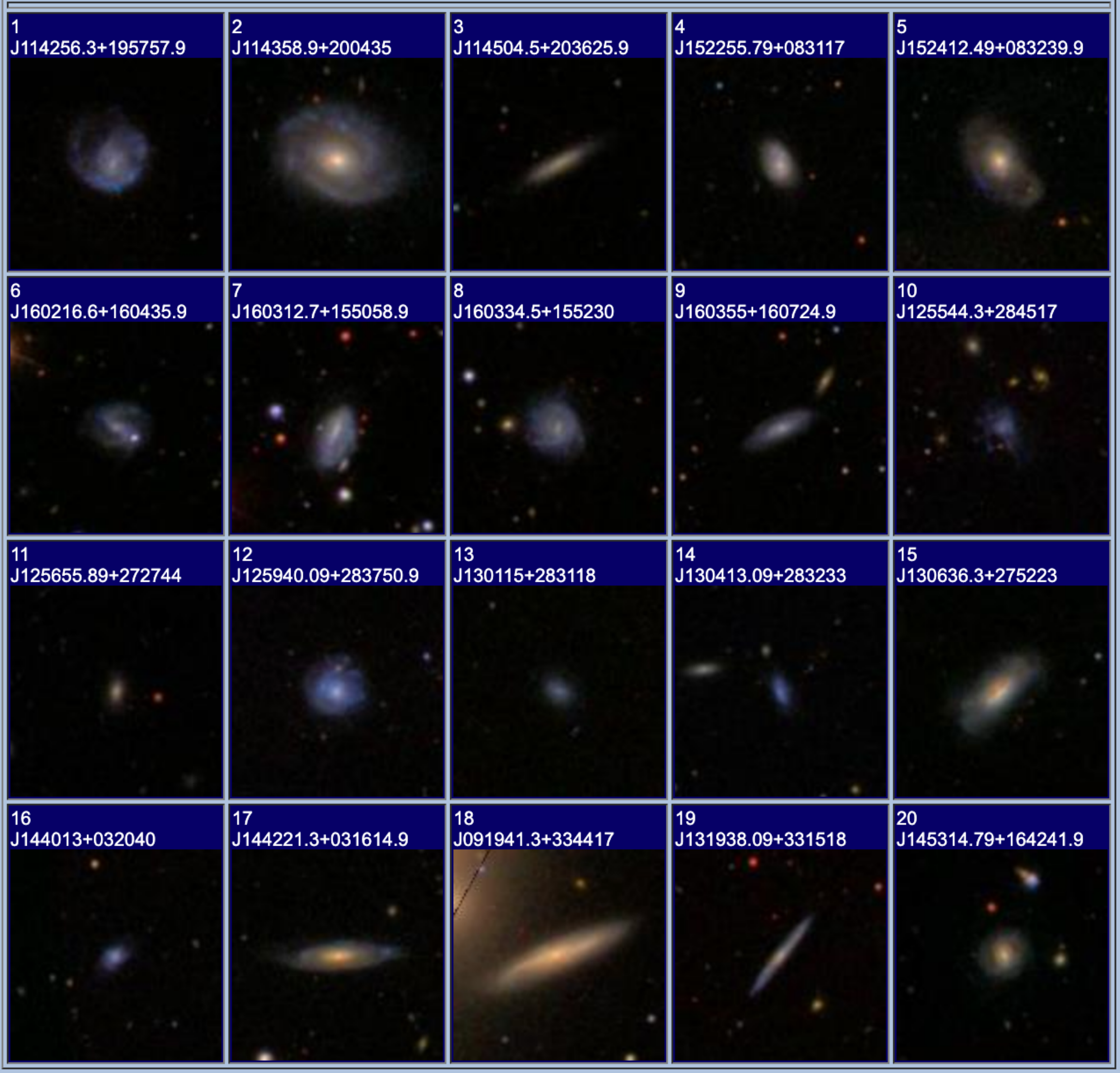}
\caption{ SDSS false-color ($g$, $r$ and $i$-band) atlas of strong-RPS galaxies in the main sample. The first 17 galaxies are in high-mass clusters and the last 3 galaxies are in low-mass clusters. All images are 100 arcsec in width. }
\label{fig:S_RPS}
\end{figure*}

\begin{figure*} 
\centering
\includegraphics[width=9cm]{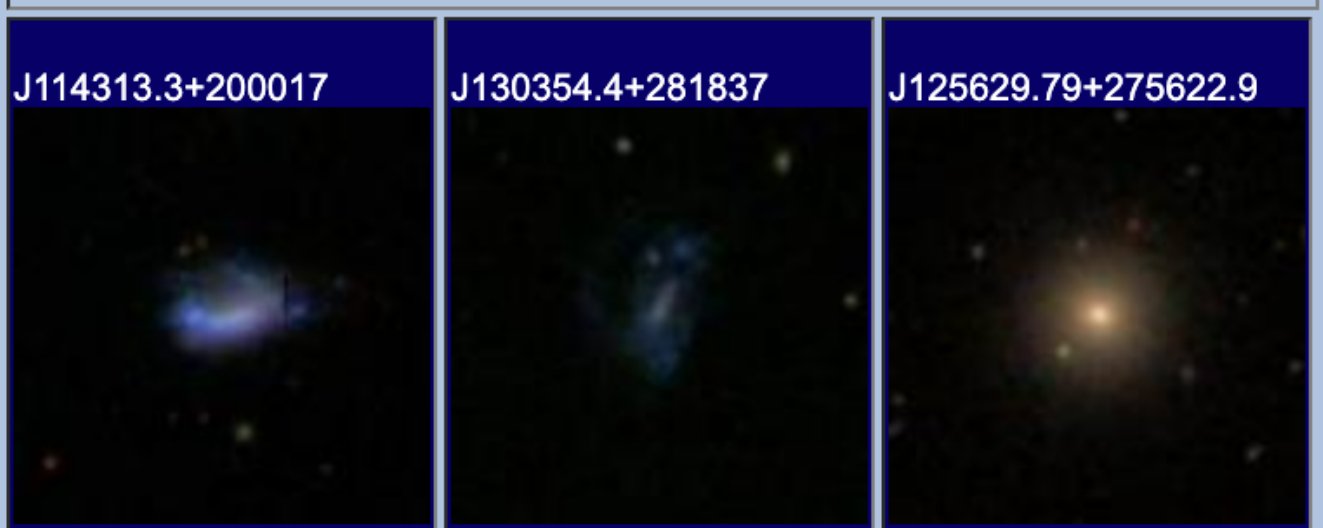}
\caption{ SDSS false-color ($g$, $r$ and $i$-band) atlas of galaxies in high-mass clusters which satisfy the selection criteria of the strong-RPS type, but were excluded from the main sample due to insufficient control galaxies. All images are 100 arcsec in width. }
\label{fig:S_RPS_nocontrol}
\end{figure*}

We present in Fig.~\ref{fig:S_RPS} the SDSS (DR7) false-color atlas for all strong-RPS galaxies in the main sample. We also present in Fig.~\ref{fig:S_RPS_nocontrol} the false-color atlas for the 3 galaxies which can be classified as the strong-RPS type, but excluded for not having enough control galaxies. We find asymmetric distribution of blue light indicative of perturbation on the gas in a few extreme galaxies, e.g. galaxy with $ID=1$ in Fig.~\ref{fig:S_RPS}, and the first galaxy in Fig.~\ref{fig:S_RPS_nocontrol}. In general it is hard to see morphological features indicative of RPS from these images, for the ($g$, $r$, and $i$-band) optical light is dominated by the old stars which are little influenced by the RPS. The clearest characteristic in the morphology is that these galaxies are disc-dominated (except for the galaxy in the right panel of Fig.~\ref{fig:S_RPS_nocontrol}, see discussion in Sec.~\ref{sec:results}), consistent with the sample selection for $\hi$-rich galaxies.

\subsection{The ALFALFA HI spectrum}
\label{sec:appendix_spec}
\begin{figure*} 
\centering
\includegraphics[width=18cm]{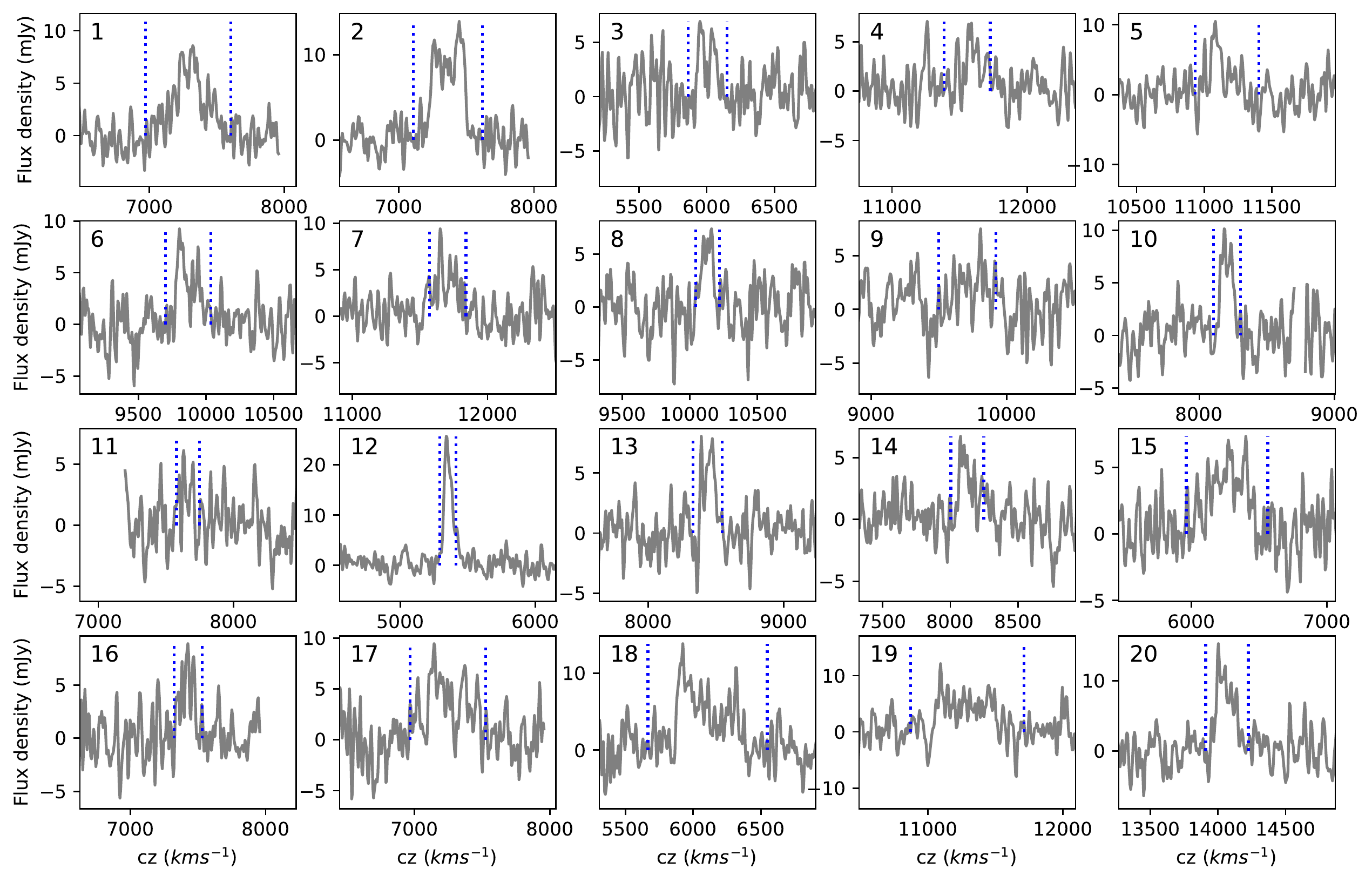}
\caption{ ALFALFA $\hi$ spectrum of strong-RPS galaxies in the main sample. Each spectrum is centered on the radial velocity of the galaxy and has a width of 1600 $km~s^{-1}$. We have smoothed each spectrum using a hanning kernel with a width of 5 channels. The measurements of $W_{50}$ (width at half the peak flux density) are marked as the dotted blue lines. The ID in each panel is consistent as that in Fig.~\ref{fig:S_RPS}.  }
\label{fig:spec_SRPS}
\end{figure*}

We present in Fig.~\ref{fig:spec_SRPS} the $\hi$ spectrum from ALFALFA for all strong-RPS galaxies in the main sample. There are some galaxies with strongly lopisided $\hi$ emission lines (e.g. $ID=$14, 18, 20), but there are also symmetric ones (e.g. $ID=$1, 2, 3). Whether the asymmetry of an integral $\hi$ line shape reflects the perturbation of the disk depends on the galactic inclination and the signal-to-noise ratio of the spectrum \citep{Watts20}. Calibration against a sample of solidly confirmed RPS galaxies (either in observation or in simulation) may be needed in the future regarding whether$\slash$how information about RPS can be drawn from the integral spectral shape.

\subsection{Cross-matching to nearby galaxies known with RPS tails}
\label{sec:appendix_LV}
Among the 17 strong-RPS galaxies in the main sample, 4 galaxies indeed show RPS features in various image observations ($\hi$, $H\alpha$, and $u$-band optical images, \citealt{Gavazzi01, Scott10, Yagi17, Roberts20}). Particularly, CGCG 097-073 in Abell 1367 cluster (No. 1 in Fig.~\ref{fig:S_RPS}) has long, extended ionized ($H\alpha$) gas tail due to ram pressure \citep{Gavazzi01, Yagi17}. In addition, the asymmetric $\hi$ distribution of CGCG 097-073 suggests that this galaxy is undergoing strong RPS \citep{Scott10}. Three galaxies (GMP 5821, GMP 3253, GMP 597, corresponding to No. 10, 12, and 14 in Fig.~\ref{fig:S_RPS}) in Coma cluster are visually classified into potential RPS galaxies, based on CFHT $u$-band images. The $u$-band images of three galaxies show RPS features such as asymmetric star formation and tails \citep{Roberts20}.

Among the three strong-RPS galaxies in high-mass clusters excluded due to insufficient number of control galaxies (Fig.~\ref{fig:S_RPS_nocontrol}), J114313.3+200017 (CGCG 097-079 in Abell 1367 cluster) also has an extended ionized gas tail \citep{Gavazzi01, Yagi17}, as CGCG 097-073 does. Its $\hi$ peak is off from the optical center toward the ionized gas tail \citep{Scott10}.  J130354.4+281837 (GMP 713 in Coma cluster) has been reported as an RPS candidate by visual inspection of the $u$-band image \citep{Roberts20}. However, J125629.79+275622.9 (NGC 4817), an early-type galaxy, has no signature of the RPS effect reported so far. 

We note that, not all the galaxies displaying RPS tails in the literature are identified by our method as strong-RPS galaxies, mostly due to our selection criteria. For example, source J125628.57+271728.6, J125809.23+284230.9 and J125839.95+264534.3 in the Coma cluster are identified as RPS candidates by \citet{Roberts20}. The former two galaxies also show marginally asymmetric $\hi$ disks \citep{Bravo-Alfaro01}. 
They were excluded from our main sample because they have $R_{\rm HI}<R_{25}$, and should be at a relatively later stage of gas depletion. Another known example galaxy of this type is NGC 4569 in the Virgo cluster \citep{Boselli18, Chung09}.

The last galaxy among the three literature candidates was identified as a weak-RPS galaxy by our method.

\section{Test the classification method with VIVA HI images}
We take the $\hi$ interferometric data of Virgo cluster galaxies observed in the VIVA project \citep{Chung09}. We derive $\Sigma_{\rm HI}$ radial profiles and $R_{\rm HI}$ from these $\hi$ images.
We also use the SDSS photometric measurements from the Extended Virgo Cluster Catalog \citep{Kim14}, to derive $M_*$ and the optical scale-length $R_d$. 
$M_*$ is estimated from the $r$ band luminosity and the $g-r$ color, using the formula from \citet{Zibetti09}.
$R_d$ is estimated as $R_{50}$/1.678, assuming an exponential radial distribution.

We select galaxies with $M_*<10^{11}~M_{\odot}$, $R_{\rm HI}>R_{25}$, and $R_{25}>2b_{maj}$ as for the main sample, but we do not apply the selection criteria on $\Delta v_{rad}$ or $d_{proj}$. It results in 16 galaxies. We note that due to the lower distance of Virgo, VIVA reaches lower $M_{\rm HI}$ and $M_*$ limits than our main sample. We use the same set of cluster parameters as in \citet{Yoon17} to derive ram pressure and the PSD. 

\begin{figure} 
\centering
\includegraphics[width=9cm]{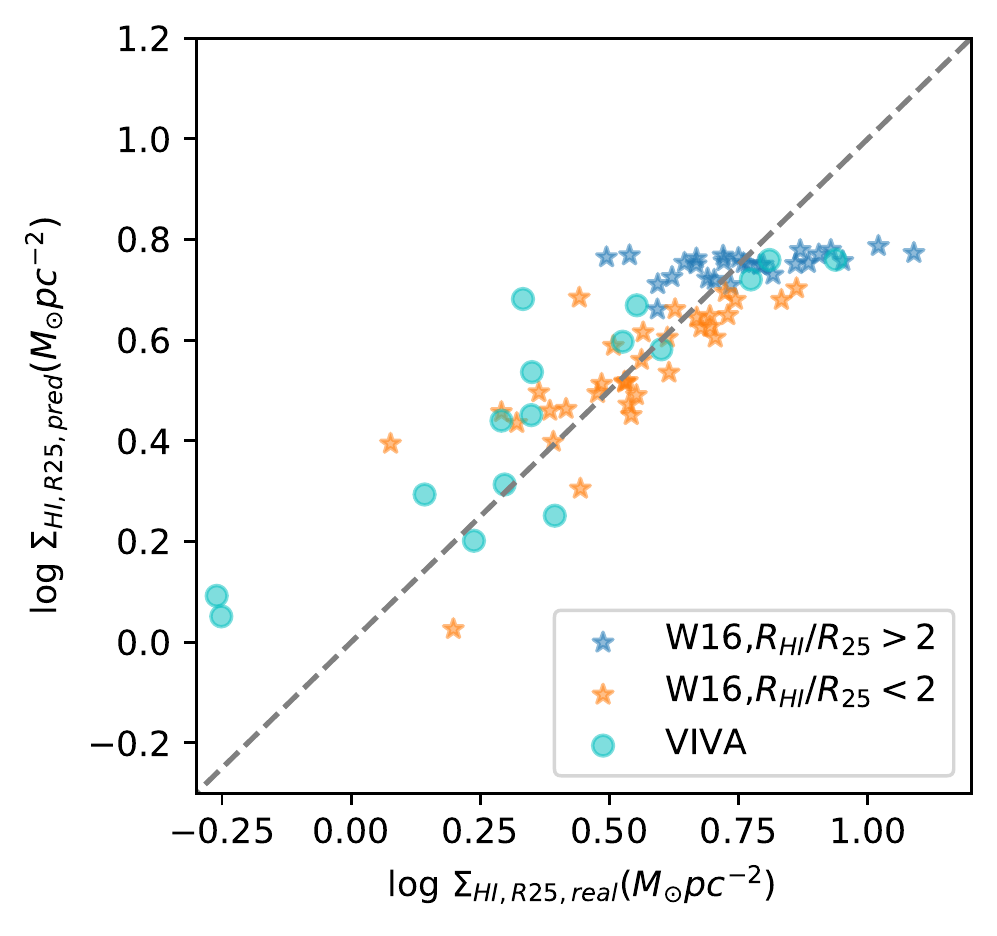}
\caption{ Comparing the predicted and measured $\Sigma_{\rm HI, R25}$. The stars mark galaxies selected from \citet[][W16]{Wang16}, by requiring $R_{\rm HI}>R_{25}$, $R_{25}>2b_{maj}$, and availability of $\Sigma_{\rm HI}$ radial profiles. The W16 sample is further divided by whether the value of $R_{\rm HI}/R_{25}$ is $>2$ (in blue) or $<2$ (in orange). T The VIVA galaxies are marked by cyan dots. The grey dashed line is the line of equivalence.  }
\label{fig:SigmaHI_R25}
\end{figure}

 A figure of comparing the predicted to the real $R_{\rm HI}$ of the VIVA galaxies can be found in \citet[][W16]{Wang16}. The two types of $R_{\rm HI}$ are close to each other, with a median offset of 0.05$\pm0.09$ dex.
 We present in Fig.~\ref{fig:SigmaHI_R25} the comparison between real and predicted $\Sigma_{\rm HI, R25}$, along with that of normal late-type galaxies from the sample of W16. Despite the good correlation (with Pearson correlation coefficients of 0.8 and 0.89 for the W16 sample and VIVA sample respectively) and relatively small scatter in the offsets between the measured and predicted $\Sigma_{\rm HI, R25}$ (0.12 dex and 0.16 dex for the W16 sample and VIVA sample respectively), we notice a saturation of the predicted $\Sigma_{\rm HI, R25}$ at $\sim6.3 M_{\odot}~pc^{-2}$. Such a saturation typically happens when the galaxy is highly $\hi$-rich and the ratio of $R_{\rm HI}/R_{25}>2$, so that $R_{25}$ reaches the inner non-exponential part of the median $\Sigma_{\rm HI}$ profile which has a much larger scatter than the outer part (see Fig. 2 of W16, and Fig. 1 of \citealt{Wang20}). The saturation tends to  under-estimate $\Sigma_{\rm HI, R25}$ and thus the anchor force at $R_{25}$, potentially rendering galaxies to be mistakenly identified into the strong-RPS type. Luckily, because $\hi$-richness tends to drop while RPS strength grows, this type of $\hi$-rich galaxies are rare within massive clusters when the ram pressure starts to work, and none of our strong-RPS galaxies have $R_{\rm HI}/R_{25}>2$. So this problem of saturation in the predicted $\Sigma_{\rm HI, R25}$ does not significantly affect the estimate of RPS strengths or the results in this paper.

\begin{figure} 
\centering
\includegraphics[width=9cm]{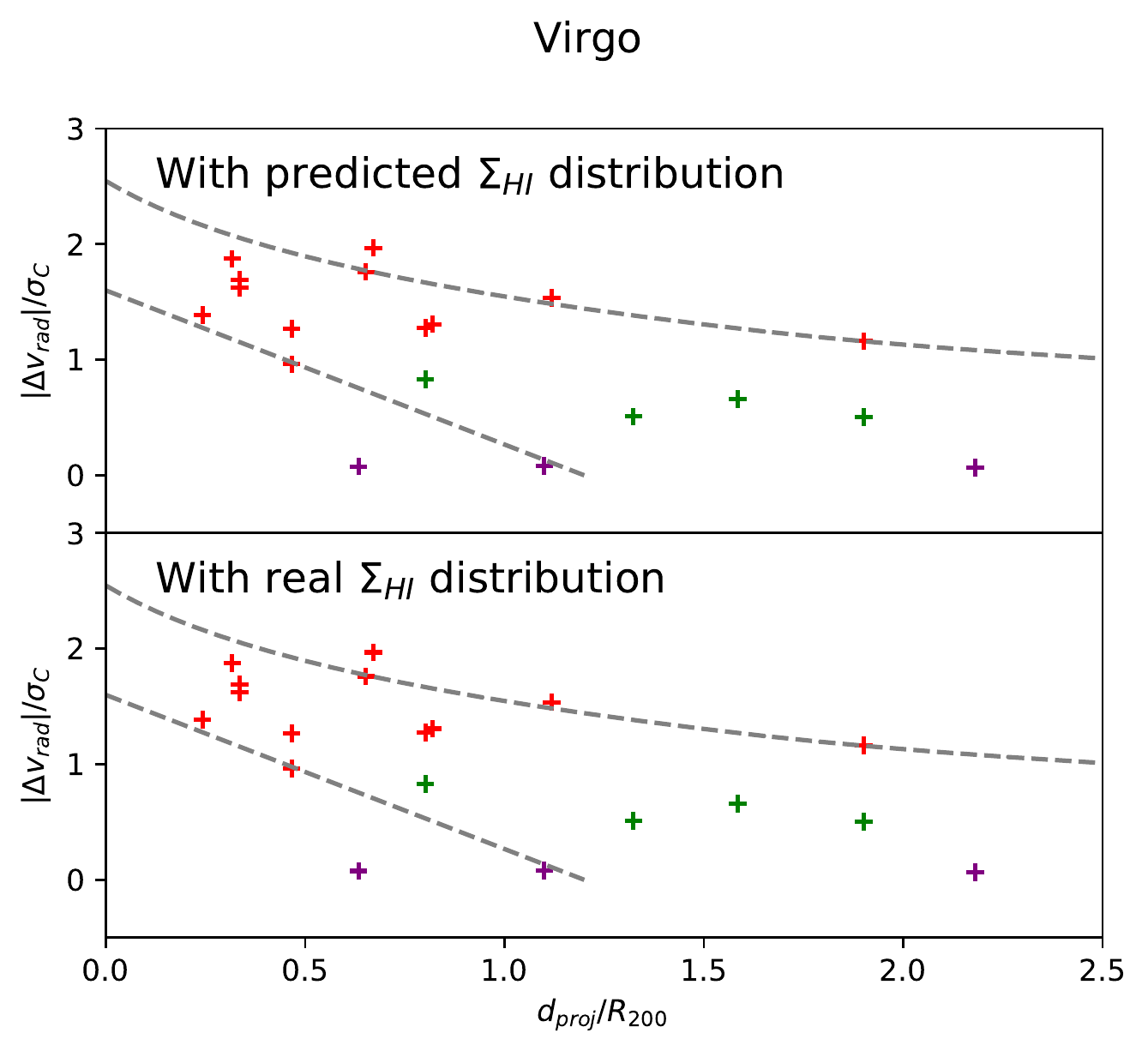}
\caption{ VIVA galaxies in the PSD. Similar as the bottom panels of Fig.~\ref{fig:PSD}, but for galaxies from the VIVA sample. The classification of galaxies into strong (red), weak (green) and no (purple)-RPS types is based on predicted and observed $\Sigma_{\rm HI}$ radial distributions in the top and bottom panel, respectively. }
\label{fig:PSD_VIVA}
\end{figure}

We use the directly measured and predicted $\Sigma_{\rm HI}$ radial distributions respectively, and make two sets of classifications that divide galaxies into the strong, weak and no-RPS types. We compare the two sets of classifications on the PSD of the Virgo cluster (Fig.~\ref{fig:PSD_VIVA}), and find them to be identical. It strongly supports our classification based on predicted $\Sigma_{\rm HI}$. We also see from Fig.~\ref{fig:PSD_VIVA} that most of the galaxies (12/19) are classified into the strong-RPS type, consistent with the overally perturbed $\hi$ morphologies reported before for the sample \citep{Chung09}. 
 As discussed in Sec. 3.1 and 5.3, our classification method is only applicable to statistical samples, but inappropriate to discussion of individual sources. We refer the readers to \citet{Yoon17} for a comprehensive discussion about the statistical correspondence between the perturbed $\hi$ morphologies and the PSD positions for the VIVA galaxies.

\end{document}